\documentclass[article]{aa}
\usepackage{epsfig}
\usepackage{ulem}
\usepackage{graphics}
\usepackage{amssymb}

\def\roph{L1688-W}
\def\deg{$^{\circ}$}
\def\h{$^{\rm h}$}
\def\m{$^{\rm m}$}
\def\s{$^{\rm s}$}

\newcommand{\lCCCHH}     {{\it l}-C$_{3}$H$_{2}$}
\newcommand{\cCCCHH}     {{\it c}-C$_{3}$H$_{2}$}
\newcommand{\ccchh}     {C$_{3}$H$_{2}$}
\newcommand{\lCCCH}     {{\it l}-C$_{3}$H}
\newcommand{\cCCCH}     {{\it c}-C$_{3}$H}
\newcommand{\ccch}     {C$_{3}$H}
\newcommand{\chhhcch} {CH$_{3}$C$_{2}$H}
\newcommand{\cch} {C$_{2}$H}
\newcommand{\cccch} {C$_{4}$H}

\newcommand{\csh} {C$_{6}$H}

\newcommand{\dco} {$^{12}$CO}
\newcommand{\tco} {$^{13}$CO}
\newcommand{\cdo} {C$^{18}$O}

\newcommand{\cmt} {cm$^{-3}$}

\newcommand{\aba} {~$^{\rm (10^{21})}$}	
\newcommand{\abb} {~$^{\rm (10^{15})}$} 
\newcommand{\abc} {~$^{\rm (10^{13})}$} 
\newcommand{\abd} {~$^{\rm (10^{12})}$} 

\begin{document}

\title{Carbon budget and carbon chemistry in Photon Dominated Regions}

\author{D. Teyssier$^{1,2}$ \and D. Foss\'e$^2$ \and M. Gerin$^2$ \and J. Pety$^{2,3}$\and A. Abergel$^4$\and E. Roueff$^{5}$}               

\offprints{D.~Teyssier \\
{\it e-mail:} teyssier@sron.rug.nl}

\institute{$^{1}$Space Research Organization Netherlands, P.O.Box 800, 9700 AV Groningen, The Netherlands, under an ESA external fellowship. \\
 $^{2}$Laboratoire d'Etude du Rayonnement et de la Mati\`ere, UMR 8112, CNRS, Ecole Normale Sup{\'e}rieure et Observatoire de Paris, 24 rue Lhomond, 75231 Paris cedex 05, France\\
 $^{3}$Institut de Radioastronomie Millim\'etrique, 300 rue de la Piscine, F-38406, St Martin d'H\`eres, France\\
 $^{4}$Institut d'Astrophysique Spatiale, Universit\'e Paris-Sud, B\^at. 121, 91405 Orsay Cedex, France\\
 $^{5}$LUTH, UMR8102 du CNRS, Observatoire de Paris, Place J. Janssen, 92195 Meudon Cedex, France
}
\authorrunning{D. Teyssier et al.}
\titlerunning{Carbon budget and carbon chemistry in PDRs}
\date{Received 17 October 2003/ Accepted 9 December 2003}

\abstract{
We present a study of small carbon chains and rings in Photon Dominated Regions (PDRs) performed at millimetre wavelengths. Our sample consists of the Horsehead nebula (B33), the $\rho$\,Oph L1688 cloud interface, and the cometary-shaped cloud IC63. Using the IRAM 30-m telescope, the SEST and the Effelsberg 100-m telescope at Effelsberg., we mapped the emission of \cch, $c$-C$_3$H$_2$ and C$_4$H, and searched for heavy hydrocarbons such as $c$-C$_3$H, $l$-C$_3$H, $l$-C$_3$H$_2$, $l$-C$_4$H$_2$ and C$_6$H. The large scale maps show that small hydrocarbons are present until the edge of all PDRs, which is surprising as they are expected to be easily destroyed by UV radiation. Their spatial distribution reasonably agrees with the aromatic emission mapped in mid-IR wavelength bands. \cch~and $c$-C$_3$H$_2$ correlate remarkably well, a trend already reported in the diffuse ISM (Lucas and Liszt 2000). Their abundances relative to H$_2$ are relatively high and comparable to the ones derived in dark clouds such as L134N or TMC-1, known as efficient carbon factories. The heavier species are however only detected in the Horsehead nebula at a position coincident with the aromatic emission peak around 7\,$\mu$m. In particular, we report the first detection of \csh~in a PDR. We have run steady-state PDR models using several gas-phase chemical networks (UMIST95 and the New Standard Model) and conclude that both networks fail in reproducing the high abundances of some of these hydrocarbons by an order of magnitude. The high abundance of hydrocarbons in the PDR may suggest that the photo-erosion of UV-irradiated large carbonaceous compounds could efficiently feed the ISM with small carbon clusters or molecules. This new production mechanism of carbon chains and rings could overcome their destruction by the UV radiation field. Dedicated theoretical and laboratory measurements are required in order to understand and implement these additional chemical routes.
}
\maketitle

\section{Introduction}

Carbon is the fourth most abundant element in the interstellar medium (ISM), and also the most versatile for building molecules. Carbon chemistry can therefore be considered as the core of interstellar chemistry. Of the nearly 130 molecules now observed in various sources, about 75\% have at least one carbon atom, while one fourth are hydrocarbons. Moreover, the heaviest and most complex molecules are organic molecules with carbon. This statistic does not take into account the Polycyclic Aromatic Hydrocarbons (PAHs), nor the Diffuse Infrared Band carriers (DIB) which are most likely large organic molecules (Herbig 1995). Carbon, neutral or ionized, is also one of the main reactants in interstellar chemistry networks, due the large number of organic molecules, but also to its versatility allowing it to participate to numerous chemical reactions at any temperature, from the very cold dense cores, to warm and hot gas. Therefore understanding the carbon chemistry is of major importance for astrochemistry, and for star formation.

A large fraction of the chemical reactions in ISM networks involve the numerous hydrocarbons present in the ISM. Such molecules were first reported in circumstellar shells, where the chemistry is particularly favorable to their formation (\cccch~-- Gu\'elin et al. 1978, C$_5$H -- Cernicharo et al. 1986), but also in molecular dark clouds, where they are efficiently shielded from the interstellar radiation (\ccchh~-- Thaddeus 1981 et al., \cch -- Gottlieb et al. 1983). Since then, hydrocarbons of increasingly {\bf greater} number of carbons have been reported in several objects (e.g. C$_6$H -- Gu\'elin et al. 1987, C$_8$H -- Dickens et al. 2001). In the diffuse gas, molecules like CH, CH$^+$ and CN are known since the 40's, and carbon clusters (C$_2$, C$_3$) are now almost routinely detected in the visible towards bright stars (Maier et al. 2001, Roueff et al. 2002, Oka et al. 2003) as well as in the far-IR (e.g. Cernicharo et al. 2000). Using radio telescopes, Lucas \& Liszt (2000) have also shown that \cch~and $c$-\ccchh~are ubiquitous in diffuse gas, confirming previous work by Cox et al. (1989).

A natural question therefore arises: if carbon chains are present in the diffuse ISM, what does happen in the Photon-Dominated Regions (hereafter PDRs) ? As in diffuse clouds, the chemical processes are dominated by the radiation but the gas is denser. The radiation field can also be constrained (intensity and direction) and, with the new data obtained by ISO, an accurate picture of the mid-IR emission due to the Aromatic Infrared Band (AIB) carriers at these cloud interfaces has emerged. Knowledge about the distribution of carbon chain and rings in PDRs is however yet scarce and limited to the works of Fuente et al. (1993, 2003), who reported the observation of \cch~at some positions in NGC7023, the Orion Bar and NGC7027, and Ungerechts et al. (1997), who presented maps of the Orion Bar in the millimetre transitions of \cch~and $cyclic$-\ccchh. In this paper we extend the study of hydrocarbons in PDRs through an extensive inventory of carbon chains and rings of up to six carbon atoms. Our aims are to get a better view of their distribution on large scale with respect to other standard tracers, and to derive their contribution to the total carbon budget in comparison with values measured in other regions.

The paper is organised as follows. We present in Sect.~\ref{presentation} the three sources studied here, while the observations are described in Sect.~\ref{obs}. We then analyse the spatial distribution of the mapped hydrocarbons (Sect.~\ref{spatial}) and derive the molecular column densities of the observed species (Sect.~\ref{coldens}). These results are compiled in the form of a carbon budget and compared to similar observations reported in dark clouds and in the diffuse ISM (Sects.~\ref{abundances} and~\ref{comparisons}). We finally compare the inferred abundances to numerical PDR models (Sect.~\ref{model}) and present our conclusions in Sect.~\ref{conclusion}. 

\section{Presentation of the sources}
\label{presentation}

 \begin{figure}
 \begin{center}
 \centerline{\epsfig{file=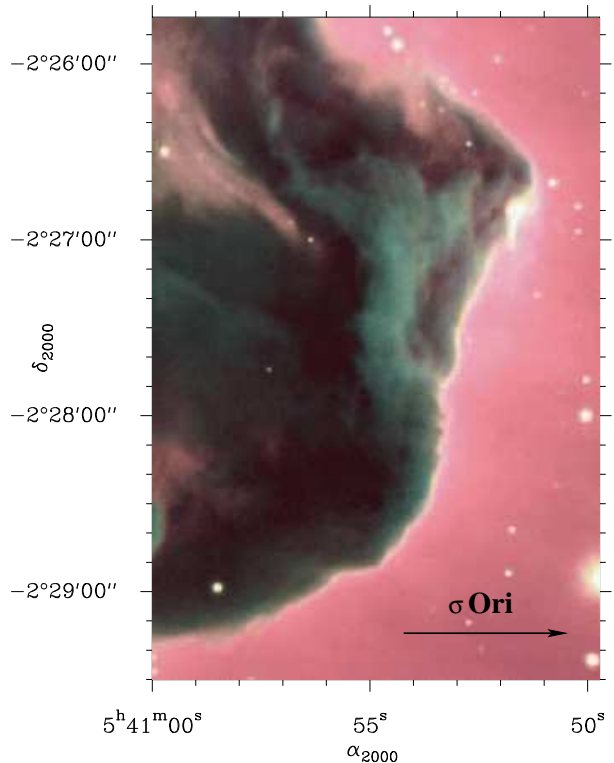, width=8.cm, angle = 0}}
 \caption{Composite colour image of the Horsehead nebula PDR obtained by the VLT in the B, V and R bands. The arrow indicates the direction of the illuminating star $\sigma$Ori. Image courtesy of ESO.}
 \label{hh fig}
 \end{center}
 \end{figure}


 \begin{figure}
 \begin{center}
 \centerline{\epsfig{file=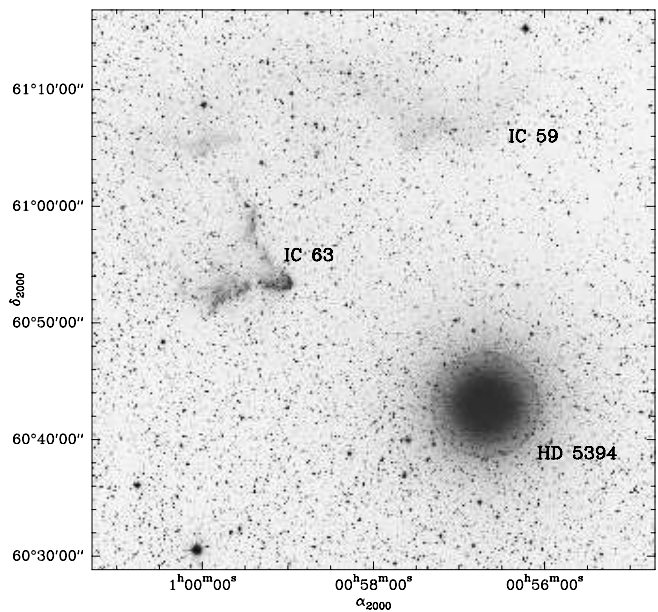, width=8.cm, angle = 270}}
 \caption{Visible image from the Digital Sky Survey in the area of IC63. The cometary shape originates from the strong radiation field emitted by the close-by HD5394 stars 20\arcmin~South-West.}
 \label{ic63 fig}
 \end{center}
 \end{figure}


 \begin{figure}
 \begin{center}
 \centerline{\epsfig{file=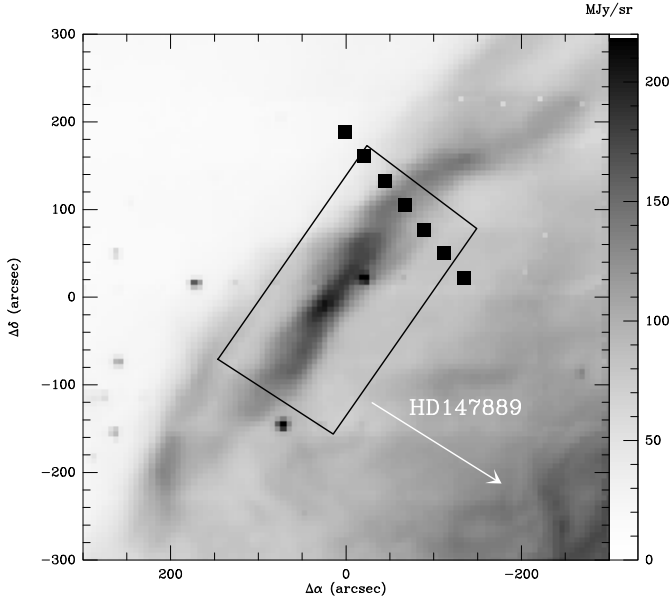, width=8.8cm, angle = 0}}
 \caption{Close-up of the western border of L1688 mapped by ISOCAM around 7\,$\mu$m (Abergel et al. 1996). The illuminating star HD147889 direction is indicated by an arrow. The black squares indicate the positions of the cut where molecular data have been collected. The tilted rectangle delimitates the area mapped in $^{13}$CO and C$^{18}$O. Offsets are given with respect to $\alpha_{2000}$=16\h25\m58\s, $\delta_{2000}= -24$\deg21\arcmin00\arcsec.}
 \label{roph fig}
 \end{center}
 \end{figure}


 \begin{figure*}
 \begin{center}
 \centerline{\epsfig{file=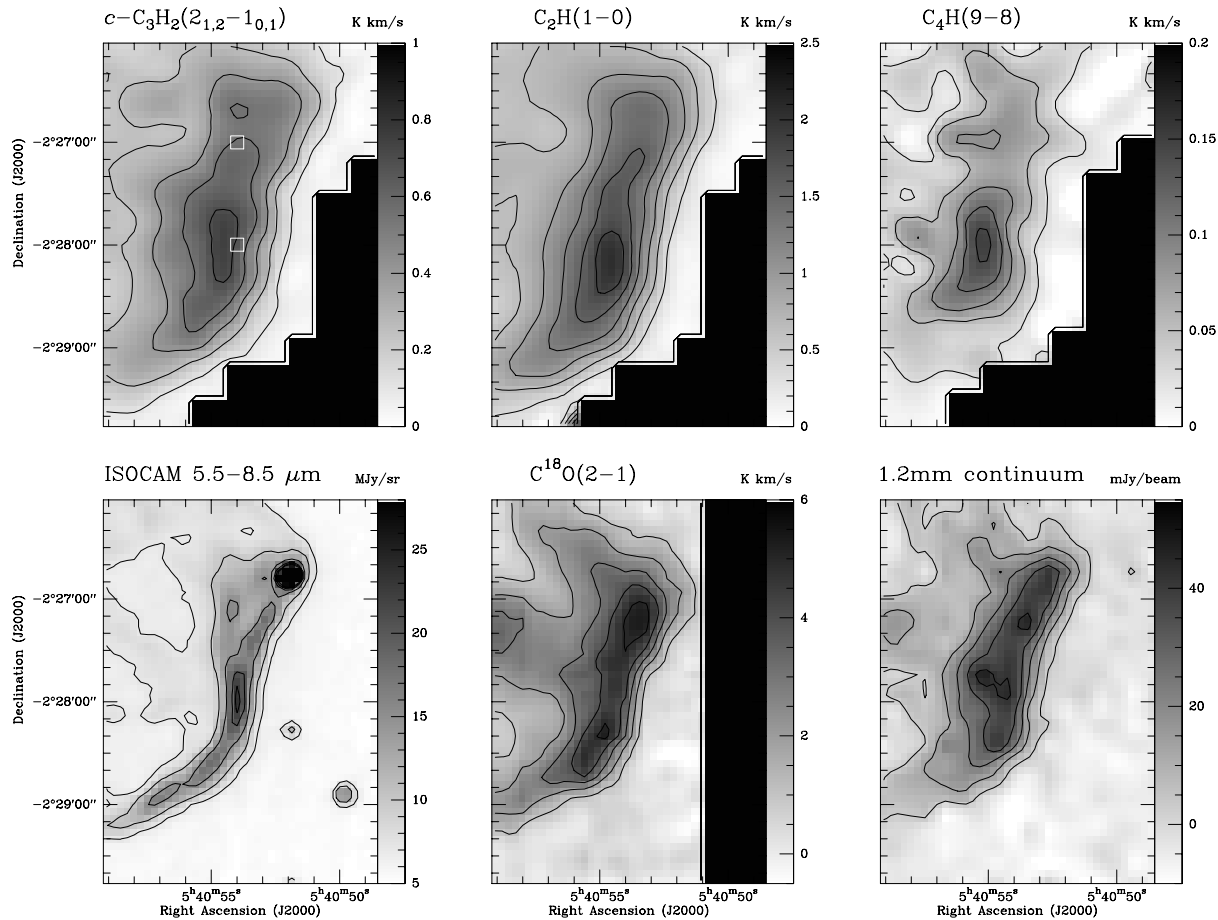, height=16.cm, angle = 270}}
 \caption{Integrated emission maps ($T_{\rm A}^*$, K\,km/s) of the molecular emission in the Horsehead nebula PDR for several species. Contours are: {\it c}-C$_3$H$_2$(2$_{1,2}$-1$_{0,1}$) 0.1 to 0.7 by 0.15, C$_2$H(1--0) 0.3 to 2.1 by 0.3, C$_4$H(9-8) 0.025 to 0.2 by 0.03, C$^{18}$O(2--1) 1 to 6 by 1. Also displayed is the continuum emission between 5 and 8.5\,$\mu$m (aromatic emission) as mapped by ISO (from Abergel et al. 2002, contours are 10 to 25 by 5 MJy/sr), and at 1.2\,mm (contours are 3 to 53 by 10 mJy/11\arcsec~beam) mapped by the MAMBO 117-channel bolometer. The white squares indicated the position labelled as CO-peak (top) and IR-peak (bottom).}
 \label{hh maps}
 \end{center}
 \end{figure*}


\subsection{The Horsehead nebula}
\label{presentation HH}

In the visible, the Horsehead nebula (B33, Barnard 1919) appears as a dark patch seen in silhouette against the ionized H$\alpha$ emission emanating from the IC434 H{\sc ii} region. The large-scale maps reported by e.g. Maddalena et al. (1986, CO(2--1)) or Lada et al. (1991, CS) show its connection to the L1630 cloud, which belongs to the Orion B molecular complex (see e.g. Abergel et al. 2002 for an overview of the region).
In the scenario proposed by Reipurth \& Bouchet (1984), the Horsehead nebula is {\bf interpreted as} an early Bok globule emerging from its parent cloud via the eroding incident radiation field emitted by the close-by O9.5V star $\sigma$Ori. The Horsehead nebula thus corresponds to a condensation resisting the illumination and it presents on its western border a PDR whose detailed shape is displayed on Fig.~\ref{hh fig}. Assuming a distance of 400\,pc (Anthony-Twarog 1982), and considering a projected distance to $\sigma$Ori of 0.5$^\circ$, the incident radiation field is of order $G_0=100$ (Zhou et al. 1993, Abergel et al. 2003). This can be seen as relatively weak compared to other well-known PDRs of the solar neighborhood (e.g. the Orion Bar, NGC7023). This PDR was mapped by ISO (6\arcsec~resolution) in filters around 7 and 15\,$\mu$m (Abergel et al. 2002, Abergel et al. 2003). It reveals a very thin filament (10\arcsec, or $\sim$0.02\,pc at 400\,pc) whose narrow size suggests that the PDR is seen perfectly edge-on (see also Fig.~\ref{hh maps}).

Little attention was paid to this object until recently. In the CO and isotope maps of Kramer et al. (1996), the nebula is heavily diluted in the 2\arcmin~beam and only appears as a small intrusion of L1630 into the H{\sc ii} region. These large-scale data already indicated mean densities of order $3\times10^{4}$\,cm$^{-3}$ (Zhou et al. 1993) and column densities compatibles with $A_{\rm V}\sim 3$ (Kramer et al. 1996). Observations at higher spatial resolution (10-20\arcsec) were only reported very recently. Pound et al. (2003) presented BIMA CO(1--0) and discussed in details the possible origin and evolution of the dark cloud. They reveal a complex velocity structure but their data are likely affected by non-negligible optical depth effects. Abergel et al. (2003) report IRAM 30-m maps obtained in the ($J$=1--0) and ($J$=2--1) transitions of CO, $^{13}$CO and C$^{18}$O. They infer densities compatible with the values reported at larger scale, and show clear selective photo-dissociation of the considered species at the PDR edge.

\subsection{IC63}
\label{presentation ic63}

IC63 is a reflexion nebula associated to the B0.5IVpe variable star $\gamma$ Cas (HD5394, $G_0 = 1100$) located 230$\pm$70\,pc from us (Vakili et al. 1984). This strong radiation field has created a PDR at the western border of the cometary-shaped molecular cloud reported by Jansen et al. (1994). IC63 is associated to another reflexion nebula, namely IC59, located at a projected distance of 20\arcmin~(or 1.3\,pc), as is illustrated on Fig.~\ref{ic63 fig}. 
According to Blouin et al. (1997), IC59 would be lying in the background of the IC63-HD5394 pair. Jansen et al. (1994), on the other hand, claim that the illuminating star would be closer to the observer, so that IC63 is partially seen face-on.

The molecular component appears elongated and seems well aligned with the direction of $\gamma$ Cas. Jansen et al. (1994) report that the CO and CS emissions are confined into a cometary cloud of size $\sim$\,1\arcmin$\times$2\arcmin, and assume an even smaller area of 30\arcsec$\times$20\arcsec~for all other species than CO. This has particular consequences in terms of beam dilution. These authors derive $T_{\rm kin} = 50\pm 10$\,K and n$_{\rm H_2} = 5\pm2 \times10^{4}$\,cm$^{-3}$, with a possible density gradient along the cloud major axis. The inferred H$_2$ column densities translate into $A_{\rm V} = 6.3\pm2.5$\,mag. Finally, an analysis of the carbon budget in this source indicates that the total gas phase carbon abundance  (5.4$\times$10$^{-5}$) is only 13\% of the solar one, which suggests that the bulk of the carbon is in solid phase and might be located on PAHs or larger grains.

IC63 has also been mapped with ISOCAM in the CVF mode allowing imaging spectroscopy between 5 and 16.5\,$\mu$m (see Foss\'e et al. 2000). Fig.~\ref{ic63 maps} shows the emission observed around the 6.2\,$\mu$m AIB feature, which corresponds to a stretching mode of the C-C bond. The aromatic emission follows the PDR border and exhibit several peaks just behind the PDR and inside the molecular tail.

\subsection{The $\rho$ Ophiuchi L1688 cloud}
\label{presentation roph}

The Ophiuchi molecular complex is a close-by star formation region (distance 135$\pm$15\,pc, Habart et al. 2003). While its so-called ``North'' area is considered inactive, the West counterpart contains two main massive cores, L1688 and L1689, heated on large scales by the Sco-OB2 association. We concentrate here on the L1688 complex, which hosts numerous pre-stellar cores (e.g. Motte et al. 1998), as well as hundreds of YSOs (Greene \& Young 1992, Bontemps et al. 2001), very likely responsible for the strong luminosity in the infrared. 

ISOCAM observations revealed bright interlaced filaments of width $\sim 0.03$\,pc coinciding with the western border of the cloud (Fig.~\ref{roph fig}, see also Abergel et al. 1996). It corresponds to a PDR (hereafter called \roph) illuminated by the B2V star HD147889. According to Abergel et al. (1999), this star is located in the centre of a spherical cavity exhibiting irregular edges. The filamentary structure is interpreted as the edge-on regions of this cavity.
Assuming that the star and PDRs are in a common plane perpendicular to the line of sight, Habart et al. (2003) infer an incident radiation field of $G_0 = 400$. Based on the AIB and the H$_2$(1--0) S(1) line emissions, these authors predict a density plateau of n$_{\rm H}=4\times 10^4$\,cm$^{-3}$ inside the cloud, followed by a decrease of the form n$_{\rm H}(r)\propto r^{2.5}$ towards the radiating star, where $r$ is the radial dimension across the PDR. This drop occurs at $\Delta\alpha\sim-100$\arcsec~on the cut indicated in Fig.~\ref{roph fig}.

On large scales, the mid-IR emission shape is well reproduced by the CO and isotopes maps obtained by Lada \& Wilking (1980) and Wilking\& Lada (1983). Their data revealed a complex velocity structure and, in particular, significant self-absorption in the $^{13}$CO species. On these scales, $^{12}$CO data are compatible with $A_{\rm V}$ in the range 50-100\,mag.

\begin{table}
\begin{center}
\begin{tabular}{l c c c } 
\hline \hline  
Molecule	&Transition		&Frequency	&HPBW		\\
		&    			&(GHz)  	&(arcsec)   	\\ \hline \hline
$^{12}$CO	& J=2--1 		& 230.538000    & 11 		\\
$^{13}$CO 	& J=2--1 		& 220.398686 	& 22/44		\\
C$^{18}$O	& J=1--0 		& 109.782160    & 22/44		\\
          	& J=2--1 		& 219.560357 	& 11 		\\
CS		& J=2--1		& 97.980950	& 24		\\
C$_2$H  	&N=1--0, J=3/2--1/2 	&   		&     		\\
        	&F=2--1 		&87.316925 	&28/56 		\\
        	&F=1--0 		&87.328624 	&28/56 		\\
        	&N=1--0, J=1/2--1/2 	&   		&		\\
        	&F=1--1 		&87.402004 	&28/56 		\\
        	&F=0--1 		&87.407165 	&28/56 		\\
		&N=3--2, J=7/2--5/2	&	&		\\
		&F=4--3			&262.004260	&28$^{(a)}$	\\
		&F=3--2			&262.006480	&28$^{(a)}$	\\
		&N=3--2, J=5/2--3/2 	&		&		\\
		&F=3--2			&262.064990	&28$^{(a)}$	\\
		&F=2--1			&262.067460 	&28$^{(a)}$	\\
\cCCCH  	&N=2$_{12}$--1$_{11}$, J=5/2--3/2 & 	&     		\\
        	&F=3--2 		&91.494349 	&27/54 		\\
        	&F=2--1 		&91.497608 	&27/54 		\\
        	&N=2$_{12}$--1$_{11}$, J=3/2--1/2 &	&		\\
        	&F=1--0 		&91.692752 	&27/54		\\
        	&F=2--1 		&91.699471 	&27/54		\\
\lCCCH  	&J=9/2--7/2             &     		&		\\
        	&F=5--4 (f)		&97.995166	&25/50		\\
        	&F=4--3 (f)		&97.995913	&25/50		\\
        	&F=5--4 (e)		&98.011611	&25/50		\\
        	&F=4--3 (e)		&98.012524	&25/50		\\
\cCCCHH 	&1$_{1,0}$--1$_{0,1}$	&18.343145	&54$^{(b)}$	\\
        	&2$_{1,2}$--1$_{0,1}$	&85.338898	&28/56		\\
\lCCCHH		&1$_{0,1}$--0$_{0,0}$	&20.792590	&48$^{(b)}$	\\
        	&3$_{3,0}$--2$_{2,1}$	&216.27875	&11		\\
\cccch  	&N=9--8, J=19/2--17/2	&		&		\\
        	&F=9--8 		&85.634006 	&28/56		\\
        	&F=10--9		&85.634017 	&28/56		\\
        	&N=9--8, J=17/2--15/2	&		&		\\
        	&F=8--7 		&85.672581 	&28/56		\\
        	&F=8--8 		&85.672583 	&28/56		\\
$l$-\cccch$_2$	&$11_{0,11}-10_{0,10}$	&98.245016	&24		\\
\chhhcch	&J=5--4                 &     		&		\\
        	&K=2       		&85.450730	&28		\\
        	&K=1       		&85.455622	&28		\\
        	&K=0       		&85.457272	&28		\\
\csh	  	&$^2\Pi_{3/2}$ J=59/2--57/2 e	&81.777893	&28		\\
    	  	&$^2\Pi_{3/2}$ J=59/2--57/2 f	&81.801247	&28		\\
HN$^{13}$C	&J=1--0			&		&		\\
		&F=0--1			&87.090735	&27		\\
		&F=2--1			&87.090859	&27		\\
		&F=1--1			&87.090942 	&27		\\
HC$_3$N		&J=9--8			&81.881468	&28		\\
\hline
\end{tabular}
\caption{Line parameters for the all the species and data reported in this paper. When two spatial resolutions are indicated, we refer to the HBPW's corresponding to the IRAM 30-m and the SEST telescopes respectively. $^{(a)}$: observed at the CSO. $^{(b)}$ Observed at the Effelsberg 100-m telescope. See Sect.~\ref{obs} for details.}
\label{tab1}
\end{center} 
\end{table} 


\section{Observations}
\label{obs}

The data presented in this paper have been gathered between September 1999 and May 2002 at the IRAM 30-m, the CSO, the SEST and the Effelsberg 100-m\footnote{The 100-m telescope is operated by the MPIfR (Max-Planck-Institut f\"ur Radioastronomie)} telescopes. They {\bf are} concentrated on a series of molecules including CO and its main isotopomers, as well as a collection of small carbon chains and rings. The details on the considered lines are compiled in Tab.~\ref{tab1}. We give here some additional information for each source. In the following the prefixes {\it c-} and {\it l-} will respectively refer to the {\it cyclic} and {\it linear} isomers of the considered species.

\subsection{Horsehead nebula}
\label{obs HH}

The Horsehead nebula PDR was mapped at the 30-m telescope in the ($J$=1--0) and ($J$=2--1)~transitions of C$^{18}$O, as well as in {\it c}-C$_3$H$_2$(2$_{1,2}$--1$_{0,1}$), C$_2$H(1--0) and C$_4$H(9--8) (Fig.~\ref{hh maps}). System temperatures at 3\,mm were in the range 100-120\,K. Some weaker lines were probed at dedicated positions (Figs.~\ref{hh points} and~\ref{ic63 points}) in CS, {\it c}-C$_3$H(2$_{1,2}$--1$_{1,1}$), {\it l}-C$_3$H(9/2--7/2), {\it l}-C$_3$H$_2$(3$_{3,0}$--2$_{2,1}$), {\it l}-C$_4$H$_2$($11_{0,11}-10_{0,10}$), HN$^{13}$C(1--0), CH$_3$CCH(5$_{\rm k}$--4$_{\rm k}$) and \csh(59/2-57/2). Except from the on-the-fly maps of C$^{18}$O, all spectra were obtained in the Frequency Switching mode using an autocorrelator spectrometer providing 80\,kHz resolution elements.
Additional data were obtained in C$_2$H(3--2) at the CSO (AOS spectrometer, $T_{\rm sys}\sim 800$\,K), and in {\it l}-C$_3$H$_2$(1$_{0,1}$--0$_{0,0}$) and {\it c}-C$_3$H$_2$(1$_{1,0}$--1$_{0,1}$) at the Effelsberg 100-m telescope ($T_{\rm sys}\sim 90$\,K). Moreover, a 1.2\,mm continuum emission map was obtained at the 30-m using the new 117-channel MAMBO bolometer (11\arcsec~resolution). Using a fast-mapping mode (Teyssier \& Sievers 1999), we could cover areas of 8.5\arcmin$\times$7.5\arcmin~in about 30\,minutes, with different chopping throws and at different hour angles, reducing efficiently some mapping artifacts inherent to the EKH inversion technique applied here (Emerson et al. 1979). Only part of this map is displayed here (Fig.~\ref{hh maps}) and further {\bf results from} these data will be discussed in a forthcoming paper.

The data were first calibrated to the $T_{\rm A}^*$ scale using the so-called chopper wheel method (Penzias \& Burrus 1973). The final adopted scale however depends on the source size. In the particular case of the Horsehead nebula, we applied the correction factor introduced in Abergel et al. (2003) taking into account the error beam of the 30-m telescope. At 3\,mm, we assumed hydrocarbon emission areas similar to the C$^{18}$O one.

\subsection{IC63}
\label{obs ic63}

\begin{figure}
 \begin{center}
 \centerline{\epsfig{file=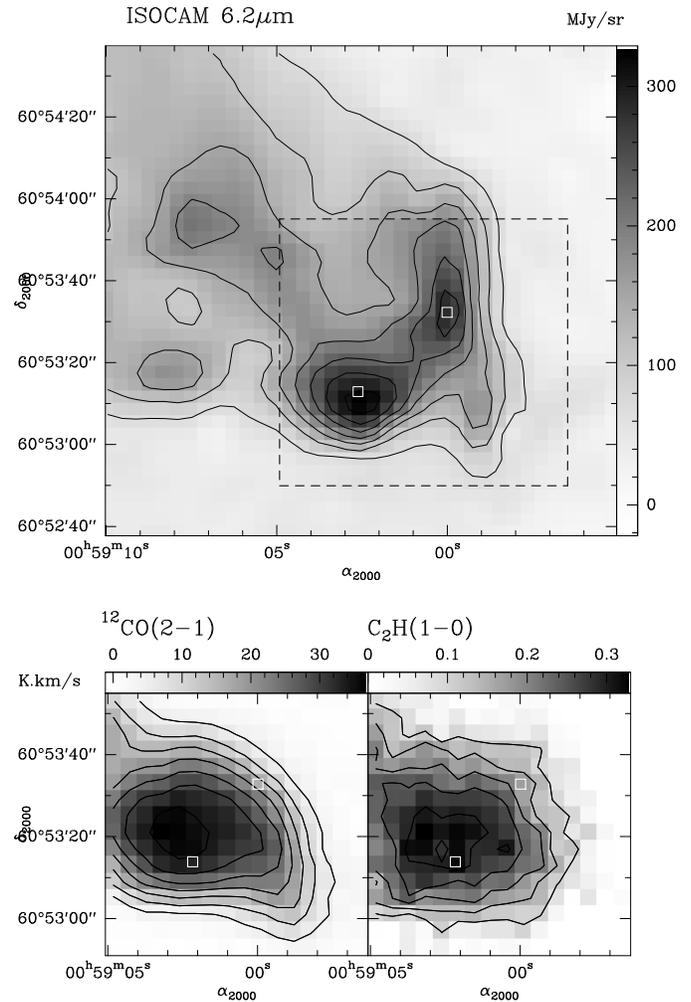, width=8.8cm, angle = 0}}
 \caption{Maps of IC63. {\it Top:} Integrated emission of the ISOCAM CVF spectra between 5.98 and 6.54\,$\mu$m after subtraction of a 1-order baseline. Contours are 70 to 310 by 40 MJy/sr. The dashed box indicates the area covered by the molecular millimetre data displayed below. {\it Bottom:} Integrated emission maps ($T_{\rm A}^*$, K\,km/s) of $^{12}$CO(2--1) (contours 5 to 40 by 5) and C$_2$H(1--0) (contours 0.10 to 0.35 by 0.05). The white squares indicate the position of the IR peaks.}
 \label{ic63 maps}
 \end{center}
 \end{figure}

IC63 was mapped at the 30-m in CO(2--1) and C$_2$H(1--0) using the on-the-fly mode (Fig.~\ref{ic63 maps}). Orthogonal coverages were optimally combined using the PLAIT algorithm by Emerson \& Graeve (1988). In addition to C$^{18}$O(2--1), the same lines as in the Horsehead were probed along a cut crossing the PDR (see Fig.~\ref{ic63 points}). All other observing parameters were similar to the Horsehead ones. Dedicated positions were also observed at the 100-m in {\it c}-C$_3$H$_2$(1$_{1,0}$--1$_{0,1}$).

After calibrating to the $T_{\rm A}^*$ scale, the lack of large-scale information in the probed molecules motivated us to adopt the widely-used main beam brightness temperature $T_{\rm mb}$ assuming beam efficiencies at the IRAM 30-m of $\eta_{\rm b}$ (= $B_{\rm eff}/F_{\rm eff}$ in the IRAM nomenclature) = 0.79 and 0.49 at 3\,mm and 1.3\,mm respectively. An additional correction was considered to account for the source dilution in the beam (see the description by Jansen et al. 1994) applying deconvolved sizes as inferred in Sect.~\ref{result ic63}.

\subsection{\roph}
\label{obs roph}

\begin{figure}
 \begin{center}
 \centerline{\epsfig{file=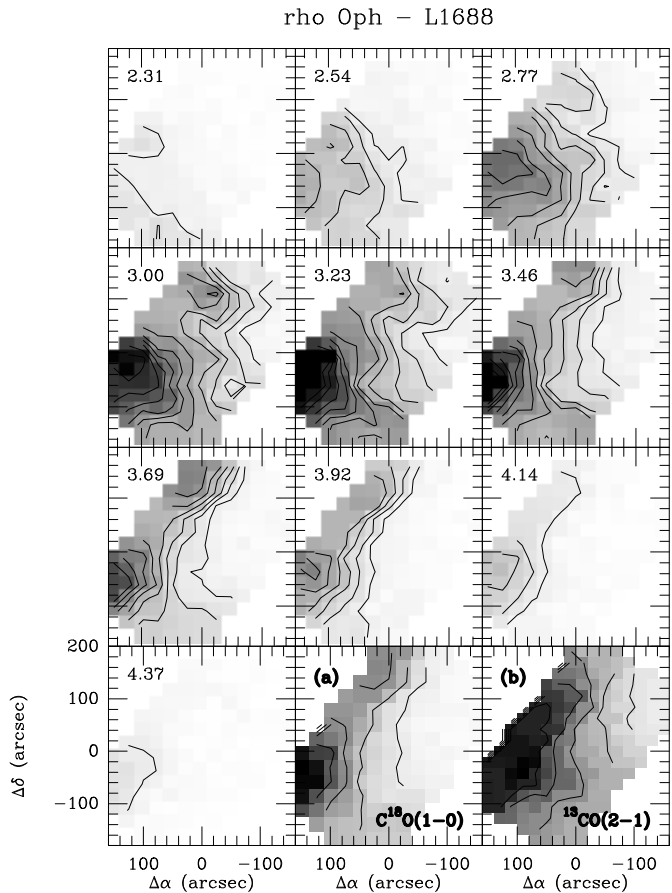, width=8.8cm, angle =0}}
 \caption{Channel maps of the C$^{18}$O(1--0) emission (K\,km/s, $T_{\rm A}^*$) in the \roph~PDR, as well as integrated intensity maps of C$^{18}$O(1--0) and $^{13}$CO(2--1) ({\bf (a)} and {\bf (b)}) in the velocity interval 3.4-4.2\,km/s, corresponding to the emission associated to the PDR as seen by ISO. The velocities of each channel map is indicated in the upper left corner. Offsets are given with respect to $\alpha_{2000}$=16\h25\m58\s, $\delta_{2000}= -24$\deg21\arcmin00\arcsec.}
 \label{roph maps}
 \end{center}
 \end{figure}

The \roph~cloud was observed at the SEST in July 2000. We mapped the $^{13}$CO(2--1) and C$^{18}$O(1--0) line emissions in an area indicated on Fig.~\ref{roph fig}. The area we have covered is just at the edge of the large-scale map of Wilking \& Lada (1983). This figure also shows the position of the cut performed in Frequency Switching for some of the small hydrocarbon species listed above. Fig.~\ref{roph maps} displays the integrated intensity maps of the two CO isotopomers. The corresponding observing parameters can be found in Tab.~\ref{tab1}.

As for IC63, the data were finally scaled to main beam brightness temperature, assuming SEST beam efficiencies of $\eta_{\rm b}=0.78$ and 0.56 at 3\,mm and 1.3\,mm respectively.

 \begin{figure}
 \begin{center}
 \centerline{\epsfig{file=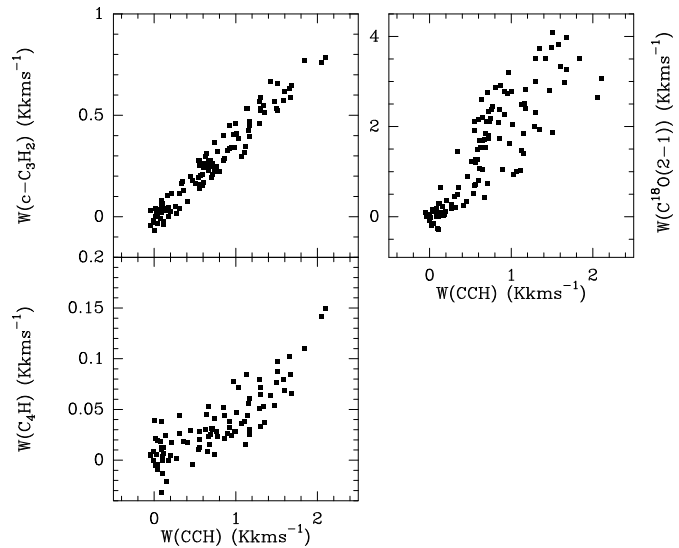, height=8.8cm, angle = 270}}
 \caption{Scatter diagrams in the Horsehead nebula. The scatter around the zero emission points is indicative of the noise in the data.}
 \label{hh correl}
 \end{center}
 \end{figure}


\section{Results}
\label{results}

\subsection{General trends}
\label{spatial}

Despite the strong incident radiation fields ($G_0 \sim\,\,$100-1000) compared to the diffuse ISM ($G_0 \sim\,\,$1), the small carbon chains and rings are detected in all objects. At first glance, this is surprising as small hydrocarbons are expected to be rapidly destroyed by such UV radiations.   
The spatial intensity variations of all hydrocarbons measured in this study correlate well. This result was already reported by Lucas \& Liszt (2000) in the diffuse ISM for $c$-\ccchh~and \cch. Fig.~\ref{hh correl} illustrates this trend in the Horsehead nebula. 
Some additional information is given below for each of the sources.

\subsection{Horsehead nebula}
\label{result HH}

The maps obtained in the Horsehead nebula are displayed in Fig.~\ref{hh maps}. They show that the hydrocarbon large-scale distribution correlates well with the PDR structure as traced by ISO or $^{12}$CO (see maps by Abergel et al. 2003). In particular the sharp edge, although smoothed by the beams, is observed in all tracers. 

On smaller scales, different behaviours have to be mentioned. It is indeed interesting to note the different positions of the respective hydrocarbon and CO (and isotopomers) peaks. Fig.~\ref{hh correl} shows that the correlation between \cdo~and \cch~is worse than with $c$-\ccchh, though the former lines are the strongest and thus offer better signal-to-noise ratio. Moreover, the hydrocarbon peak is located very close to the peak of aromatic emission around 7\,$\mu$m (hereafter called IR peak), suggesting a chemical link between AIB carriers and the small hydrocarbons. More conclusive comparisons with the ISO mid-IR data are nevertheless limited by the 30-m spatial resolution. This has important consequences on the physico-chemical analysis along cuts crossing the PDR as, for instance, any abundances probed at the IR peak will not coincide with the peak emission of the hydrocarbons. This issue will be addressed in a companion paper presenting interferometric observations of the same molecules (resolution $\sim 5$\arcsec, Pety et al. 2003). 

\begin{figure*}
 \begin{center}
 \centerline{\epsfig{file=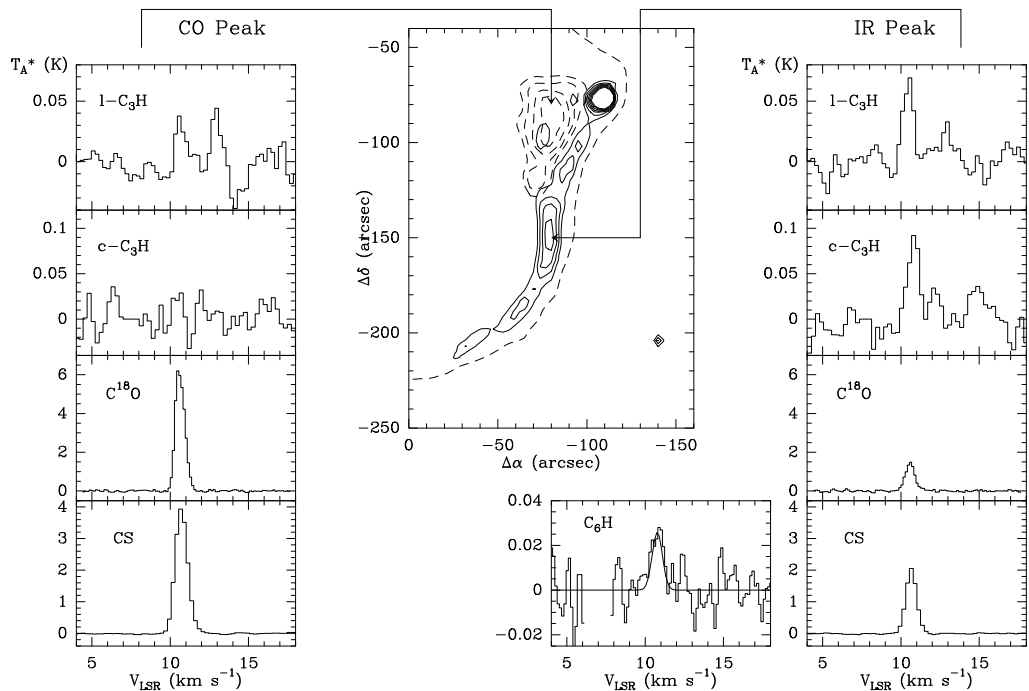, height=14.5cm, angle = 270}}
 \caption{Sample spectra of heavier hydrocarbons and other species obtained at the so-called CO peak (indicated by dashed $^{12}$CO contours, 30\% and 80 to 95\% of the peak, from Abergel et al. 2003) and IR peak (indicated by full line ISO contours 15 to 30\,MJy/sr, from Abergel et al. 2002). All temperature are in K, $T_{\rm A}^*$ scale. The same scales are used for both positions to ease the comparison between the various molecules. The missing channels in the C$_6$H spectrum were removed due to spurious features in the backend response. For this line, we also display the Gaussian fit to the line. Offsets are given with respect to $\alpha_{2000}$=05\h40\m59.0\s, $\delta_{2000}= -02$\deg25\arcmin31.4\arcsec.}
 \label{hh points}
 \end{center}
 \end{figure*}

Other species were also probed at positions of interest. They include heavier and rarer hydrocarbons, as well as some cyanopolyynes or density tracers such as CS (see Tab.~\ref{tab1}) and were chosen in order to complete our inventory of the carbon budget. In the Horsehead, these deep integrations were unfortunately performed prior to the large-scale mapping, so that the position observed at the IR peak does actually not correspond to the maximum hydrocarbon emission, as is seen on~Fig.~\ref{hh maps}. The CO peak position was inferred from the $^{12}$CO intensity map observed by Abergel et al. (2003). Although not displayed in Fig.~\ref{hh points}, HN$^{13}$C was detected at both positions, and HC$_3$N was observed at the hydrocarbon peak. $l$-\ccchh, $l$-\cccch$_2$ and \chhhcch~were not detected at either of the positions, while $c$-\ccch~and $l$-\ccch were detected only at the IR peak. All other molecules were detected to a better than 3\,$\sigma$ level. We also emphasize the detection of \csh~at the IR peak which is to our knowledge the first detection of this molecule reported in a PDR. It could however not be observed at the CO peak.

\subsection{IC63}
\label{result ic63}

The IC63 maps are shown in Fig.~\ref{ic63 maps}, as well as a cut illustrating the radial profiles of more species across the PDR border (Fig.~\ref{cut hh}). The correlation observed between the various tracers in the Horsehead is not so striking in the IC63 maps as this source is barely resolved by the 30-m observations and the large beam (28\arcsec~compared to the 6\arcsec~offered by ISO) significantly smoothes out the details at the PDR border. Smoothing the ISO data to the 30-m resolution however shows a very good agreement between the probed emissions as the two main infrared peaks get smoothed into one single maximum coincident with the molecular emission feature. On the other hand, the CO and C$_2$H peaks coincide fairly well, and are located just North to the south-most aromatic peak at 6.2\,$\mu$m (indicated on Fig.~\ref{ic63 maps}). From the observed integrated emission, a beam-deconvolved size of $\sim$~40\arcsec$\times$30\arcsec~(FWHM) can be inferred for $^{12}$CO and C$_2$H (the North-East tail is not accounted here).

\begin{figure*}
 \begin{center}
 \centerline{\epsfig{file=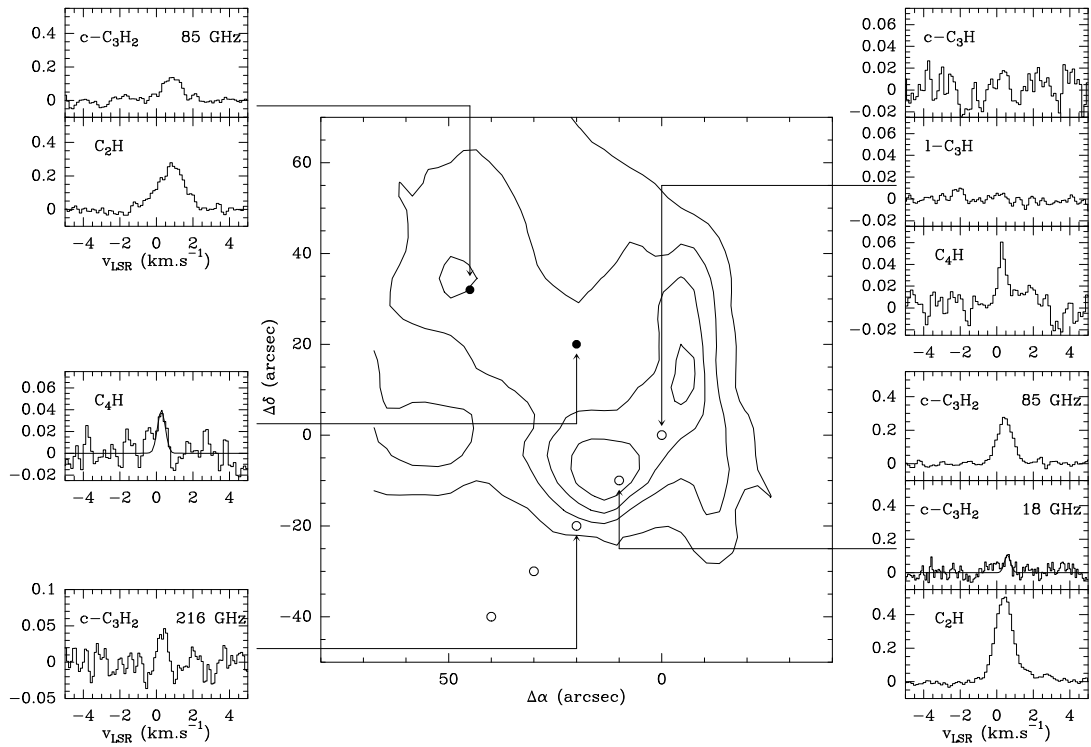, height=14.5cm, angle = 270}}
 \caption{Sample spectra of weaker hydrocarbons and other species obtained at some dedicated positions in IC63 displayed over contours of the 6.2\,$\mu$m ISO integrated emission map. It illustrates in particular the non-detection of the cyclic and linear forms of C$_3$H around 92 and 98\,GHz respectively. The probed positions are marked with filled and open circles. The open circles indicate the cuts displayed in Fig.~\ref{cut hh}. All temperature are in K, $T_{\rm A}^*$ scale. Offsets are given with respect to $\alpha_{2000}$=00\h59\m00.7\s, $\delta_{2000}= 60$\deg53\arcmin19\arcsec.}
 \label{ic63 points}
 \end{center}
 \end{figure*}

\begin{figure}
 \begin{center}
 \centerline{\epsfig{file=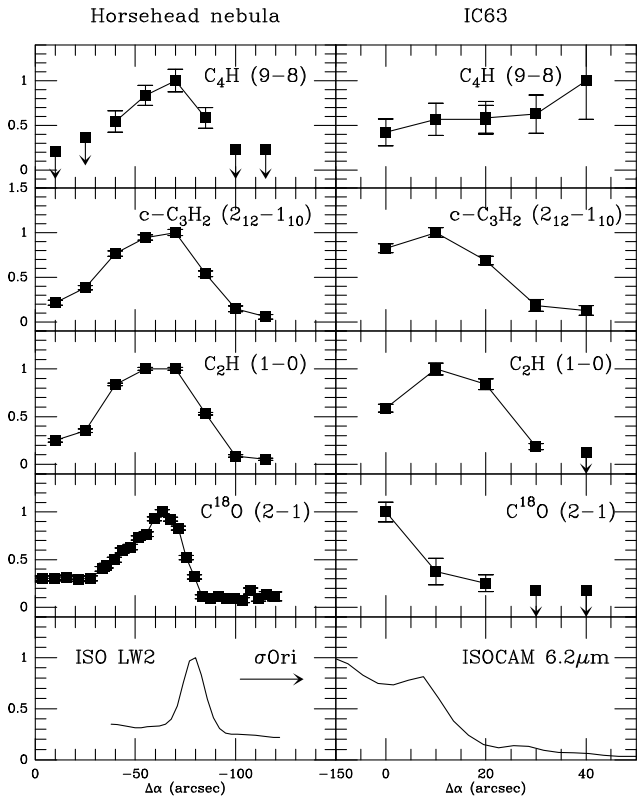, width=10.5cm, angle = 270}}
 \caption{Normalized line intensities across the PDR fronts of the Horsehead nebula and IC63. {\it Horsehead}: iso-declination cut performed through the IR-peak (see Fig.~\ref{hh maps}) at $\delta_{2000}= -2$\deg28\arcmin00\arcsec. All intensities are normalized to the peak (respectively 0.24, 0.9, 2.3 and 8.7\,K from top to bottom, and 24\,MJy/sr for the ISO LW2 profile). The illuminating source lies towards the negative offsets. {\it IC63}: cut performed along the direction indicated by open circles on Fig.~\ref{ic63 points}. All intensities are normalized to the peak (respectively 0.03, 0.27, 0.5 and 0.2\,K from top to bottom, and 310\,MJy/sr for the ISOCAM CVF profile). The illuminating source lies towards the negative offsets. All error bars are $\pm$1\,$\sigma$. Upper limits indicate non-detections at a 2\,$\sigma$ level.}
 \label{cut hh}
 \end{center}
 \end{figure}

In IC63, the deep integrations were done along a cut crossing the south-most AIB emission peak (indicated in Fig.~\ref{ic63 points}), as well as some extra positions inside the cloud tail. \cccch~and $c$-\ccchh~are in most cases detected, the latter including transitions at 18 and 85\,GHz (see Tab.~\ref{tab1}). \cch(3--2) is only marginally detected at some positions. The non-detections are similar to the ones reported in the Horsehead nebula, but include as well both isomers of \ccch, HN$^{13}$C and \csh. 

\subsection{\roph}
\label{result roph}

\begin{figure*}
 \begin{center}
 \centerline{\epsfig{file=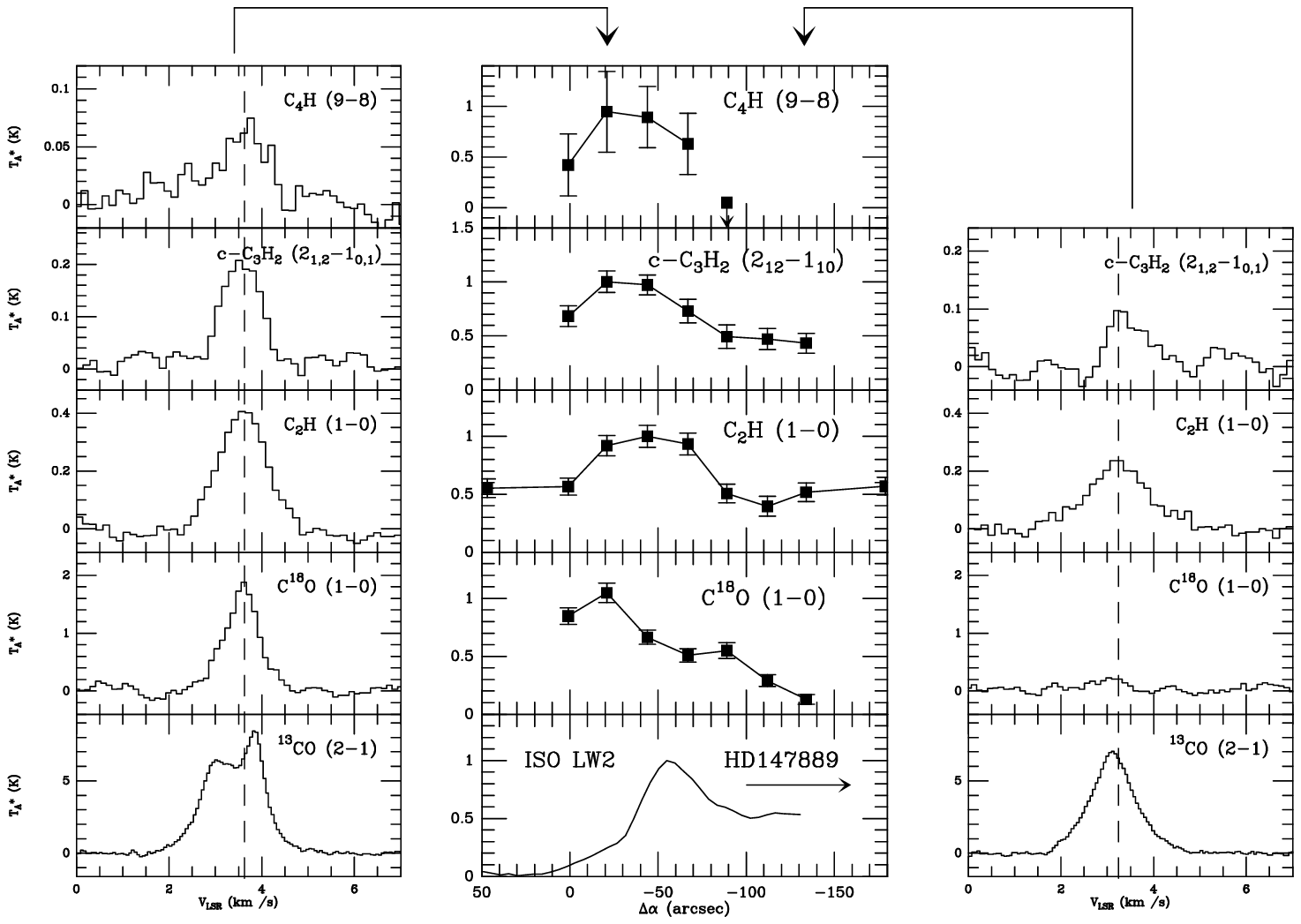, height=15.cm, angle = 270}}
 \caption{Normalized line intensities over the cut displayed on Fig.~\ref{roph fig} across the \roph~PDR. Error bars are $\pm$1\,$\sigma$. The star lies towards the negative offsets. The 0-offset refers to the most North-East point of the cut (Fig.~\ref{roph fig}). All intensities are normalized to the peak (respectively 0.08, 0.2, 0.43 and 1.6\,K from top to bottom, and 117\,MJy/sr for the ISO LW2 profile). An upper limit is indicated for the rightmost point of C$_4$H. Sample spectra at two offsets are also shown. The dashed line indicates the centroid velocity of the optically thin lines, illustrating the {\bf self}-absorption in \tco. 
All temperature are in K, $T_{\rm A}^*$ scale.}
 \label{roph points}
 \end{center}
 \end{figure*}

In \roph, the molecular cloud exhibits a complex velocity field, as was already reported by Wilking \& Lada (1983). It is illustrated on Fig.~\ref{roph maps} for the area covered in this study. We associate the PDR border as seen by ISO (Abergel et al. 1999) to the molecular emission in the velocities range 3.4 to 4.2\,km/s (see integrated intensity maps in Fig.~\ref{roph maps}). Along the cut displayed in Fig.~\ref{roph fig}, the brightest hydrocarbon emission is detected at positions towards the illuminating star where C$^{18}$O has already started to decay significantly (Fig.~\ref{roph points}). The emission at negative offsets can nevertheless be due to the particular geometry of \roph (Abergel et al. 1999), implying that the PDR is partly seen face-on at probed positions inside the cavity.

In \roph, a somewhat smaller number of species was investigated in deeper integrations. Apart from the data presented in Fig.~\ref{roph points}, observations of $c$-\ccch~and $l$-\ccch~were unfruitful down a 5\,mK (1\,$\sigma$) noise level.

\section{Analysis of the physico-chemical conditions}
\label{analysis}

We have used the available collection of species to derive the physical and chemical conditions at work in the three sources, and in particular assess a tentative carbon budget in these PDRs (see Sect.~\ref{C budget}).

\begin{table*}
\begin{center}
\begin{scriptsize}
\begin{tabular}{l | c c c| c c c| c c} 
\hline \hline  
Molecule	& \multicolumn{3}{|c|}{Horsehead nebula}  & \multicolumn{3}{c|}{IC63 cut} 			            & \multicolumn{2}{c}{\roph~cut} \\
		& CO peak &  IR peak & Hydroc. peak & (0\arcsec,0\arcsec)& (10\arcsec,--10\arcsec)	& (20\arcsec,--20\arcsec) & ($\Delta \alpha$=--21\arcsec)& ($\Delta \alpha$=--134\arcsec) \\ \hline
H$_2$\aba            & $7.5\pm 2.5$            	& $7.2\pm 2.4$		& $10.5\pm 3.5$           	& $5.0  \pm  2.0$       & $  5.0\pm  2.0$       & $5.0\pm 2.0$          & $11.4\pm 2.4$           & $2.4\pm 0.3$  \\
\cdo\abb             & $5.0\pm 0.2$		& $3.5\pm 0.2$          & $4.0\pm 0.2$			& $0.18 \pm 0.02$       & $0.05\pm0.02$                & $0.05\pm0.02$                & $4.0\pm 0.2$          & $0.4\pm0.02$  \\
($10^{-7}$)	     & {\bf 6.7}$\pm${\bf 2.2} 	& {\bf 4.9}$\pm${\bf 1.6}& {\bf 3.8}$\pm${\bf 1.3} 	& {\bf 0.36}$\pm${\bf 0.15} & {\bf 0.10}$\pm${\bf 0.06} & {\bf0.10}$\pm${\bf 0.06} & {\bf 8.0}$\pm${\bf 3.2} & {\bf 1.7}$\pm${\bf 0.2}  \\
\cch\abc             & $17.0\pm 2.0$         	& $16.0\pm 2.0$         & $22.7\pm 3.3$         	& $5.3\pm0.4$           & $8.6\pm0.6$           & $6.9\pm0.5$           & $3.0\pm 0.3$          & $2.0\pm 0.2$  \\
($10^{-8}$)	     & {\bf 2.3}$\pm${\bf 0.8} 	& {\bf 2.2}$\pm${\bf 0.8}& {\bf 2.2}$\pm${\bf 0.8} 	& {\bf 1.1}$\pm${\bf 0.4} & {\bf 2.1}$\pm${\bf 1.0} & {\bf 1.4}$\pm${\bf 0.6} & {\bf 0.6}$\pm${\bf 0.2} & {\bf 0.8}$\pm${\bf 0.1}  \\
$c$-\ccch\abd        & $< 2.1 $              	& $3.9 \pm 0.5 $	& --                    	& $< 1.2$               & --                    & --                    & --                    & --          \\
($10^{-10}$)	     & {\bf $<$ 2.8}         	& {\bf 5.4}$\pm${\bf 1.9} & --                  	& {\bf $<$ 2.4}         & --                    & --                    & --                    & --                     \\
$l$-\ccch\abd        & $< 2.3$               	& $2.1 \pm 0.7$         & --                    	& $< 5.8$               & --                    & --                    & --                    & --          \\
($10^{-10}$)	     & {\bf $<$ 3.1} 	     	& {\bf 2.9}$\pm${\bf 1.4}& --                  		& {\bf $<$ 11.6}        & --                    & --                    & --                    & --                     \\
$c$-\ccchh\abd       & $11.0\pm 3.0$         	& $9.3 \pm 0.2$         & $15.0 \pm 5.0$        	& $4.5\pm 1.0$          & $4.7\pm 1.0$          & $3.0\pm0.6$           & $3.0\pm0.6$           & $3.0\pm0.6$  \\
($10^{-9}$)	     & {\bf 1.5}$\pm${\bf 0.6} 	& {\bf 1.3}$\pm${\bf 4.3}& {\bf 1.4}$\pm${\bf 0.7} 	& {\bf 0.9}$\pm${\bf 0.4} & {\bf 0.9}$\pm${\bf 0.4} & {\bf 0.6}$\pm${\bf 0.3} & {\bf 0.6}$\pm${\bf 0.3} & {\bf 1.3}$\pm${\bf 0.3}  \\
$l$-\ccchh\abd       & $< 0.32$     	     	& $< 0.33$              & --                    	& --                    & --                    & --                    & --                    & --          \\
($10^{-11}$)	     & {\bf $<$ 4.3}         	& {\bf $<$ 4.6}         & --                    	& --                    & --                    & --                    & --                    & --                     \\
\chhhcch\abc         & $<1.1$                	& $<1.1$                & --                    	& --                    & --                    & --                    & --                    & --          \\
($10^{-9}$)	     & {\bf $<$ 1.4}         	& {\bf $<$ 1.4}         & --                    	& --                    & --                    & --                    & --                    & --                     \\
\cccch\abd           & $10.0\pm 3.0$         	& $11.0\pm 3.0$         & $31.0\pm 7.0$         	& $3.3\pm0.2$           & $3.7\pm0.4$           & $6.5\pm1.5$           & $15.0\pm 3.0$         & --          \\
($10^{-9}$)	     & {\bf 1.3}$\pm${\bf 0.6} 	& {\bf 1.5}$\pm${\bf 0.7}& {\bf 3.0}$\pm${\bf 1.2} 	& {\bf 0.6}$\pm${\bf 0.3} & {\bf 0.7}$\pm${\bf 0.3} & {\bf 1.3}$\pm${\bf 0.6} & {\bf 3.0}$\pm${\bf 1.3} & --                     \\
\csh\abd             &     --       	     	&  --                   & $1.0 \pm 0.5$         	& --                    & --                    & --                    & --                    & --          \\
($10^{-10}$)	     &    --                 	& --                    & {\bf 1.0 $\pm$0.6}    	& --                    & --                    & --                    & --                    & --                     \\
HN$^{13}$C\abd       & $1.2 \pm0.2$          	& $0.3 \pm0.1$          & --                    	& $<0.11$               & --                    & --                    & --                    & --          \\
($10^{-10}$)	     & {\bf 1.6}$\pm${\bf 0.6}	& {\bf 0.4}$\pm${\bf 0.2}& --                    	& {\bf $<$ 0.2}         & --                    & --                    & --                    & --                     \\
HC$_3$N\abd          &     --       	     	&  --                   & $0.6 \pm 0.2$           	& --                    & --                    & --                    & --                    & --          \\
($10^{-11}$)	     &    --                 	& --                    & {\bf 5.7 $\pm$2.7}      	& --                    & --                    & --                    & --                    & --                     \\
CS\abc               & $12\pm2$              	& $2.9 \pm0.5$          & --                    	& $0.6\pm0.1$           & $0.4\pm0.1$           & $0.15\pm0.03$         & --                    & --          \\
($10^{-8}$)	     & {\bf 1.6}$\pm${\bf 0.6} 	& {\bf 4.0}$\pm${\bf 1.5}& --                  		& {\bf 0.12}$\pm${\bf0.05}& {\bf 0.08}$\pm${\bf 0.04}& {\bf 0.03}$\pm${\bf 0.01}& --                    & --                    \\
\hline
\end{tabular}
\caption{Column densities and abundances relative to H$_2$ (boldface) inferred at positions of interest in the Horsehead, IC63 and \roph. 
{\bf Units for column densities are in cm$^{-2}$ and indicated in superscript for each molecule}.
Units for abundances are indicated in the first column of each corresponding row. The offsets refer to the positions indicated Figs.~\ref{roph maps} and~\ref{ic63 points}.}
\label{tab2}
\end{scriptsize}
\end{center} 
\end{table*} 


\subsection{Column densities}
\label{coldens}

As mentioned in the previous section, one should bear in mind that beam dilution effects are expected to affect the measured intensity. This is especially true in our case since we look at positions located close to sharp edges. To first-order, this mainly means that some of the column densities derived below may be under-estimated. The assumptions and methods used to compute the quantities of interest are described in details in Appendix~\ref{appendix a}. We discuss in the following some specific hypothesis on a source to source basis. The results are gathered in Tab.~\ref{tab2}.

\subsubsection{Horshead nebula}
\label{coldens HH}

From the \dco~maps of Abergel et al. (2003), we derived $T_{\rm kin} \sim 40\pm2$\,K at the PDR location. Using this temperature, the {\bf hydrogen number} densities deduced from LVG modelling are of the order of some 10$^4$\,\cmt.

Following the method described in Appendix~\ref{c2h}, we inferred \cch~excitation temperatures between 5 and 10\,K,  while $\tau_{\rm tot}$ never exceeded 1.5. We cross-checked these results using the additional N=3--2 transitions observed at some positions. The N=3--2/N=1--0 line intensity ratio are consistent with $T_{\rm ex}$ in the range 4.5--8\,K, in good agreement with the HFS outputs. The results show similar values as the one reported by Fuente et al. (1993, 2003) in NGC7023 and the Orion Bar. The constant density assumed to conduct the $c$-\ccchh~and \cccch~analysis (see Appendix~\ref{c3h2}) was taken to be 2$\times$10$^4$\,\cmt.

\subsubsection{IC63}
\label{coldens ic63}

As discussed previously, the relatively small size of the emitting area compared to the mapping beams requires a correction factor to be applied to the line intensities. We used the approach described by Jansen et al. (1994) to convert $T_{\rm mb}$ into brightness temperatures within a source filling 1000\,arcsec$^2$ for $^{12}$CO and the hydrocarbon emission, and 400\,arcsec$^2$ for the $^{13}$CO(2--1) emission. Under these assumptions, the \dco~observations indicate $T_{\rm kin} \sim 48\pm2$\,K, in excellent agreement with Jansen et al.'s finding of $50\pm10$\,K. Since only one \cdo~transition was available in this source, we made the additional assumption of a volume density n$_{\rm H_2}=5\pm2 \times10^{4}$\,cm$^{-3}$, based on the results of Jansen et al. (1994). The same density was applied to the analysis of the $c$-\ccchh~and \cccch~column densities. The inferred \cdo~column density at (0\arcsec,0\arcsec) offset is however 3 times lower than the value reported by these authors. This is likely due to the 10\arcsec~offset between their position and ours, sufficient to explain such rapid variations in regards of the source size (see Tab.~\ref{tab2}).

The \cch~analysis indicates that 80\% of the observed points exhibit $\tau_{\rm tot} < 1$ and are consistent with $T_{\rm ex}$ = 5.8--8\,K. The inferred abundances are somewhat higher than the results inferred by Jansen et al. (1994) at a position 10-20\arcsec~away from ours.

\subsubsection{\roph}
\label{coldens roph}

The CO kinetic temperatures were derived using an additional \cdo(2--1) map obtained at the CSO (Gerin, private communication). We calculated $T_{\rm kin} \sim 21\pm2$\,K in the region of interest. As already reported by Lada \& Wilking (1980), the CO lines however exhibit significant self-absorption. Despite the dilution effects mentioned above, the temperatures in this source are thus lower limits, implying upper limits for the molecular column densities.

While the \cch~HFS analysis reveal excitation temperature similar to the ones found in IC63, we used the volume density profile derived by Habart et al. (2003) to constrain the $c$-\ccchh~and \cccch~column densities. The simulations confirm the small optical depth at the probed positions.

\subsection{Fractional abundances}
\label{abundances}

\subsubsection{Systematics}

The molecular abundances with respect to H$_2$ are compiled in Tab.~\ref{tab2}. Appendix~\ref{coldens h2} describes the methods used to derive the H$_2$ column densities in each of the sources. The hydrocarbon abundances appear reasonably constant across the different objects and positions, although to a lesser extent in \roph~where abundances do not always peak at the intensity maxima. We will discuss this point later. This constancy confirms the general trend already reported in the diffuse gas (Lucas \& Liszt 2000). Although error bars remain high (essentially due to uncertainties on $N$(H$_2$)), the observed values result in average abundances of [\cch]\,$=1.9\pm0.4\times 10^{-8}$, [$c$-\ccchh]\,$=1.4\pm0.2\times 10^{-9}$ and [\cccch]\,$=1.1\pm0.2\times 10^{-9}$.

Abundance ratios between molecules show a somewhat similar trend and seem to be hierarchically distributed. On average, the following ratios are found: [\cdo]/[\cch]\,$\sim$\,25 (although with large scatter), [$c$-\ccchh]/[\cccch]\,$\sim$\,1 and [\cch]/[$c$-\ccchh]\,$\sim$\,15. The latter result is of the same order as the value reported in the diffuse gas (ratio of 28$\pm$8) by Lucas \& Liszt (2000).

\subsubsection{Horsehead nebula} 

In this source, it is interesting to note that the position where the hydrocarbon emission is the largest coincides with a peak of the H$_2$ column density as traced by the continuum emission at 1.2\,mm (Fig.~\ref{hh maps}). This trend does however not hold for species such as \cdo~(although within large error bars), CS and HN$^{13}$C, whose abundance maxima do not coincide with this H$_2$ column density peak. Note also that the fraction of the {\bf beam filled by the cloud} is smaller at the hydrocarbon peak than at the CO peak, thus lowering the intensities measured at the border. This is confirmed by interferometric observations of the PDR border which show an enhancement of the hydrocarbon abundances at the aromatic peak (Pety et al. 2003).

\subsubsection{IC63} 

In IC63, the small relative variations of abundances observed between the different offsets is very likely related to the proximity of the probed positions with respect to the beam sizes. The \cdo~abundances are significantly lower than the value reported by Jansen et al. (1994). This discrepancy arises from the large H$_2$ column densities assumed at positions probed outwards the cloud.

\subsubsection{\roph}

Firm conclusions are more difficult to draw in \roph~as the available sample remains restricted to few points. If our picture of the variations of the H$_2$ column density is correct, the highest hydrocarbon abundances are not found at their corresponding intensity peaks but closer to the illuminating star (Fig.~\ref{roph points}). Again, this can be due to the particular geometry of the PDR partly seen face-on. Still, the \cdo~abundances experience a drop much steeper than the hydrocarbons in the direction of the illuminating star.

\section{Carbon chemistry}
\label{C chem}

\subsection{Carbon budget}
\label{C budget}

The compilation of molecular abundances allows us to derive a tentative carbon budget representative of PDR conditions. As a matter of comparison, Tab.~\ref{carbon budget} also includes results obtained in the diffuse ISM (Lucas \& Liszt 2000), as well as in a typical carbon-rich dark cloud (see also Sect.~\ref{comparisons}). For this, we have co-added typical abundances for molecules with a given number of carbon atoms. Since all molecules are not observed in the same sources, this carbon budget has only a statistical meaning, and numbers cannot be very accurate. 
The contribution of carbon clusters containing up to 2-3 atoms is very similar for all three media, with a clear abundance decrease as the molecular size increases, both in PDRs and dark clouds. Heavier chains and rings (4 carbon atoms and more) are however less abundant in PDRs and diffuse clouds than in dark clouds by a factor of order 5-10. The fraction of carbon tied to more complex molecules may be only partly representative of the real situation. A lower number of molecules of that size is detected and other species still remain to be identified.
 
\subsection{Comparison to dark clouds}
\label{comparisons}

We have compared the carbon chain and ring abundances to results obtained in TMC-1 (e.g. Pratap et al. 1997) and L134N (L183, e.g. Dickens et al. 2000), two dark clouds known as efficient carbon factories. The carbon budget corresponding to the dark cloud conditions is also indicated in Tab.~\ref{carbon budget}. As for the PDRs, it remains difficult to derive accurate estimates of the total H$_2$ column densities in these objects. These estimates are indeed usually affected by strong molecular depletion (e.g. Cernicharo et al. 1987, Alves et al. 1999) and possible evolution of the grain properties in the cold and dense medium characteristic of dark clouds (e.g. Stepnik et al. 2003, Kramer et al. 2003). The abundances reported in Tab.~\ref{carbon budget} have been calculated assuming an averaged extinction of order 20\,mag, which may lie in between the actual values applying to TMC-1 and L134N. In addition we have decided to compare the fractional abundances with respect to \cch, the most abundant hydrocarbon in our sample. Assuming again an extinction of 20\,mag, the \cch~fractional abundances to H$_2$ at the TMC-1 cyanopolyyne peak (CP) and at the L134N hydrocarbon peak would however be $4\times10^{-9}$ and $8\times10^{-9}$ respectively, somewhat lower than the values we have derived in PDRs. This is also true for species such as $c$-\ccchh~or \cccch.

Tab.~\ref{tab3} compiles the abundances relative to \cch~for each of the sources at the positions of interest. The error bars have not been considered as we look here at general trends. Contrary to the abundances relative to H$_2$, the fractional abundances relative to \cch~are generally larger in the dark clouds than in our sample PDRs (we will here not address the differences between TMC-1 and L134N, see Foss\'e 2003). For all molecules but $l$-\ccch, there is at least a factor of 5 difference between the molecular fractions in the PDRs and in TMC-1. This is particularly true for abundant species such as $c$-\ccchh~and \cccch, whereas they seem to have similar contributions to the carbon budget in both media (Tab.~\ref{carbon budget}). This behaviour is very likely due to the different chemistries at work in the two media, and will be further discussed in the next section. We also note the enormous drop of [HC$_3$N]/[\cch] between TMC-1 CP and the Horsehead nebula (factor of 1000). Although we probably missed the ``cyanopolyyne'' peak position of the Horsehead, the difference is still meaningful.

\begin{table*}
\begin{center}
\begin{tabular}{l | c c c} 
\hline \hline  
Number of carbon atoms				& \multicolumn{3}{|c}{N$_{\rm Tot}$/N(H$_2$)} \\
						& PDRs		 	& Diffuse ISM			& Dark clouds				\\ \hline
1 (C$^{\rm +}$, C, CO)				& $2.6\times10^{-4}$    & $2.3\times10^{-4}$    	& $8.0\times10^{-5}$    		\\
1 (CH$^{\rm +}$, CH, CS, HCN,..)		& $7.0\times10^{-8}$    & $5.0\times10^{-8}$    	& $\leq 1.0\times10^{-7}$    		\\
2 (C$_2$, \cch, C$_2$S,..)			& $8.0\times10^{-8}$    & $8.0\times10^{-8}$    	& $1.0\times10^{-8}$    		\\
3 (C$_3$, \ccch, \ccchh, HC$_3$N,\chhhcch,..)	& $5.0\times10^{-9}$    & $4.5\times10^{-9}$    	& $1.2\times10^{-8}$    		\\
4 (C$_4$, \cccch, \cccch$_2$,..)		& $2.0\times10^{-9}$    & $1.5\times10^{-9}$    	& $1.7\times10^{-8}$    		\\
5 (C$_5$, C$_5$H, HC$_5$N,..)			& --                    & --                    	& $1.2\times10^{-9}$    		\\
6 (\csh, \csh$_2$,..)				& $1.1\times10^{-10}$   & --                    	& $4.2\times10^{-10}$    		\\
7 (HC$_7$N,..)       				& --                    & --                    	& $5.5\times10^{-10}$    		\\
8 (C$_8$H,..)					& --                    & --                    	& $1.1\times10^{-11}$    		\\
 \hline
\end{tabular}
\caption{Carbon budget derived in PDRs (this work), diffuse ISM (from Bell et al. 1983, Federman et al. 1994, Shuping et al. 1999, Lucas \& Liszt 2000, Roueff et al. 2002, Cernicharo et al. 2002 and Rachford et al. 2002) and dark clouds (from Pratap et al. 1997, Bell et al. 1998, Dickens et al. 2000, Turner et al. 2000, Foss\'e et al. 2001 and Foss\'e 2003). Note that, for a given number of carbon atoms, the molecules found in the different regions may differ (e.g. no C$_n$ carbon clusters in PDRs and dark clouds). {\bf For the diffuse ISM, we used the C/H abundance ratio of 1.4$\times$10$^{-4}$ reported by Cardelli et al. (1996) and assumed an H/H$_2$ abundance ratio of 0.6 as observed in the $\zeta$ Oph line of sight (e.g. Rachford et al. 2002).}}
\label{carbon budget}
\end{center} 
\end{table*} 


\begin{table*}
\begin{center}
\begin{tabular}{l | c c c c c} 
\hline \hline  
Molecule		& Horsehead         	& IC63 				& \roph 				& TMC-1     	& L134N \\ 
			& IR peak area		&(0\arcsec,--0\arcsec)		& ($\Delta \alpha$=--21\arcsec)		& (CP)		&	\\\hline
\cch~($10^{13}$\,cm$^{-2}$)&    16            	&		5.3		&	3.0				&   7.2~$^{(a)}$&      15~$^{(d)}$\\ \hline
$c$-\ccch             	&	0.024		&		$<$0.023	&	--				& 0.140~$^{(b)}$&  0.050~$^{(d)}$\\
$l$-\ccch              	&	0.013		&		$<$0.110	&	--				& 0.012~$^{(b)}$&  0.040~$^{(d)}$\\
$c$-\ccchh             	&	0.058		&		0.085		&	0.100				& 0.800~$^{(b)}$&  0.330~$^{(d)}$\\
$l$-\ccchh             	&	$<$0.002	&		--		&	--				& 0.030~$^{(b)}$&  0.004~$^{(d)}$\\
\chhhcch              	&	$<$0.069	&		--		&	--				&   1.4~$^{(a)}$&  0.070~$^{(d)}$\\
\cccch               	&	0.069		&		0.062		&	0.500				&   2.8~$^{(c)}$&  0.735~$^{(d)}$\\
\csh     		&	0.004		&		--		&	--				& 0.115~$^{(b)}$&  0.008~$^{(d)}$\\
HN$^{13}$C             	&	0.002		&		$<$0.002	&	--				& 0.056~$^{(a)}$&  0.054~$^{(e)}$\\
HC$_{3}$N             	&	0.003		&		--      	&	--				&   2.2~$^{(f)}$&  0.127~$^{(e)}$ \\
\hline
\end{tabular}
\caption{Fractional abundances relative to \cch. {\bf The assumed} column densities of \cch~are recalled in the first row. CP stands for Cyanopolyyne Peak. References are: $^{(a)}$~Pratap et al. 1997, $^{(b)}$~Foss\'e et al. 2001, $^{(c)}$~Gu\'elin et al. 1982, $^{(d)}$~Foss\'e 2003, $^{(e)}$~Dickens et al. 2000, $^{(f)}$~Takano et al. 1998. {\bf Since no \csh~and HC$_{3}$N measurements were available at the Horsehead IR peak, the corresponding abundances relative to \cch~are given for the hydrocarbon peak.}}
\label{tab3}
\end{center} 
\end{table*} 


 \begin{figure*} \begin{center}
 \centerline{\epsfig{file=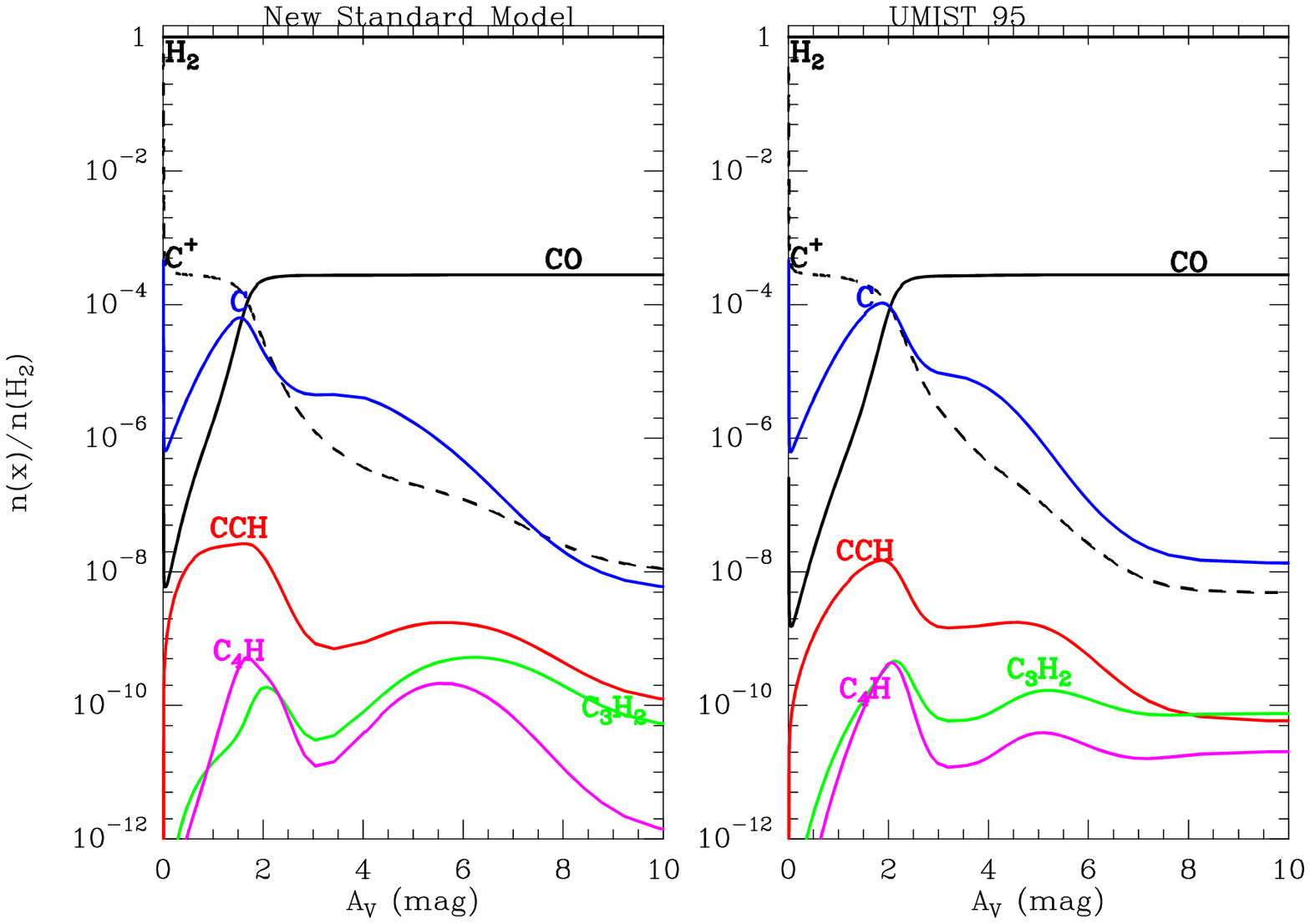, width=16.cm, angle =
 0}} \caption{Prediction for the abundances of various species derived
 from an updated Le Bourlot et al.'s PDR model, using two chemistry
 networks, namely Herbst et al.'s New Standard Model (left) and
 UMIST95 of Millar and collaborators (right). We assumed a gas density
 of n$_{\rm H}=2\times 10^4$\,cm$^{-3}$ and a radiation field
 $G_0=100$. Both models predict too low abundances for $c$-\ccchh~and
 \cccch, but also that \cch~is present in the UV illuminated gas as
 observed.}  \label{model} \end{center} \end{figure*}


\subsection{Comparison to models}
\label{model}

In this section, we compare our observed data with model
predictions. We used a plane parallel PDR model which solves
self-consistently the thermal and chemical balance of a
semi-infinite homogeneous slab (Le Bourlot et al. 1993, Le Petit et al. 2002).
We describe the results for physical conditions mimicking those of the Horsehead nebula and
the $\rho$ Oph interface, namely a gaz density of 
n$_{\rm H}$ = $2 \times 10^{4}$\,cm$^{-3}$ and a radiation field multiplied by
$G_0$ = 100 relative to the ISRF determined by Habing (1968).
Since the chemical networks have not been fully tested for
large molecules, we have chosen to use two different networks,
the New Standard Model developed by E. Herbst and collaborators 
(Terzieva \& Herbst 1998) and the
UMIST95 chemical rate file (Millar et al. 1997). 
{\bf Both chemical networks include about 400 hundred species with up to 8 carbon atoms, and 4000
chemical reactions. The various possible isomers are not
differenciated in the chemical networks.}

{\bf The photodissociation of H$_2$ and CO and the photoionization of C
are treated specifically in the PDR code via the computation of the
actual ultraviolet radiative transfer in the presence of the complex
chemistry provided by the two chemical networks. Then, the transition
of atomic to molecular hydrogen is identical in both models. However,
the  balance between neutral carbon and carbon containing species in
the intermediate range of $A_{\rm V}$ is dependent both on the
photo-processes (identical in the two networks) and the individual
chemical reactions which involve different assumptions.}

Except for the formation of molecular hydrogen, we have not
considered any reaction on grain surfaces. All molecules are therefore
formed and destroyed in the gas phase. For the external layers of the
cloud where the radiation field is the strongest, the main destruction
process is photo-dissociation, while chemical reactions are more important
 in well shielded regions. PAHs are taken into account for
their role in the gas heating, but do not participate in the
chemistry. Using two networks helps us in setting error bars on model predictions and therefore in
identifying agreement or discrepancy between our observed data and
the model predictions more securely.

The PDR model assumes steady state equilibrium for both
the thermal processes and the chemistry. Given the large
density and radiation field, this assumption should be justified since the
chemical time scales are dominated by the photo-dissociation reactions
which are very rapid in the conditions we modelled. Furthermore, 
there is no need to invoke time dependent chemistry for the 
synthesis of molecules since ionised and atomic carbon are present 
with a large abundance in the external layers (up to $A_{\rm V}$ = 4 -- 5\,mag for C), 
and can participate in a large variety of chemical
reactions for building molecules.

We show in Fig.~\ref{model} model predictions for the two chemical
networks separately. Though the transition from atomic to molecular
hydrogen is very similar for both chemical networks, the transition
from ionised carbon to neutral carbon, then to carbon monoxide, shows
small differences, with the formation of carbon monoxide occurring at
slightly smaller $A_{\rm V}$ in the NSM compared to UMIST95.

We now discuss the predictions for \cch, C$_3$H$_2$ and C$_4$H.  For
both models, the abundance of hydrocarbons presents two peaks, the
first one located near $A_{\rm V}$ = 2, and a second one near $A_{\rm
V}$ = 5. We shall compare our observed data with peak values in the
first peak, which presents the largest abundances. Given the {\bf low}
spatial resolution of the data in comparison with the expected spatial
variations, the observed abundances are lower limits to the true
abundances. For the Horsehead nebula we have good evidence from the
maps that emission lines of \cch~and related molecules peak in the
same region as the mid-IR emission and that the present data could not
resolve this sharp peak.
 
In both models, the peak abundances relative to H$_2$ 
are [\cch] $\simeq 2 \times
10^{-8}$, [C$_3$H$_2$] $\simeq 2 \times 10^{-10}$ 
and [C$_4$H] $\simeq 2 \times 10^{-10}$. The peak CO abundance is
$\simeq 2.6 \times 10^{-4}$, since all gas phase carbon is
locked into CO for $A_{\rm V}$ larger than 2 magnitudes. Compared to the observed
data, models predict similar abundances relative to H$_2$ 
for \cch, and CO ; also  C$_4$H and C$_3$H$_2$ behave in the
same  way as we observed. However, their predicted abundances relative
to H$_2$ are lower by an order of magnitude
than abundances derived from the observations. The discrepancy is
clear in both models, all the more if we consider that the peak
abundance is obtained over a fairly narrow range of $A_{\rm V}$. Since
our data are taken with a low spatial resolution, we suffer from
beam dilution as explained above.
How can we explain this discrepancy ? {\bf We suggest three possibilities which are not independent}:

\begin{itemize}
\item{The photo-dissociation rates for hydrocarbons (molecules, radicals
and ions) included in the chemical networks may be overestimated by one order of magnitude.
For most species, the values used in models are indeed best guesses with
large uncertainties (see van Dishoeck 1988 and Foss\'e 2003).}
\item{Although including thousands of reactions, chemical networks 
used for models may not list 
the relevant chemical reactions for forming carbon chains and
radicals yet. A related possibility is that some reaction rates are
incorrect. For example the role of atomic oxygen in
destroying carbon chains and cycles in  the interstellar medium is
still not settled (E. Herbst, priv. com.).
{\bf There is also the possibility that neutral neutral reactions
involving neutral carbon, CH , C2H and small hydrocarbons may be at
work and contribute to the formation of the larger hydrocarbons.
Reactions of atomic carbon with acetylene, ethylene, propyne and
allene have indeed been studied  by Chastaing et al. (2001) and shown to be
rapid.}}
\item{Another mechanism for producing  carbon chains is needed,
in addition to gas phase chemistry. An attractive possibility is the
erosion and/or destruction of PAHs and small carbon grains. Indeed
the main process constraining the smallest possible size of
PAHs is the destruction of their carbon skeleton. Various mechanisms
have been proposed (see Allain et al. 1996, Scott et al. 1997, Le Page et al. 2001, 2003, Verstraete et
al. 2001) which all lead to the production of small carbon fragments
(C$_2$H$_2$, C$_2$ or C$_3$,...). Once released in the gas phase, 
these fragments may further react
and take part in the synthesis of the carbon chains and radicals we
observed. Given the large fraction of total carbon in
solid form (PAHs can represent up to 10\% of total C) compared with
the rather small abundance of carbon chains (see Tab.~\ref{carbon budget}), the
production of carbon clusters and molecules from the PAHs is a plausible mechanism. {\bf Such a mechanism was also mentioned by Fuente et al. (2003) to explain the high hydrocarbon abundances observed in NGC7023 and the Orion Bar.} 
} 
\end{itemize}

\section{Conclusions}
\label{conclusion}

We have conducted an extensive search for small hydrocarbon chains and rings in three Photon-Dominated Regions. The map obtained in the millimetre transitions of \cch, $c$-\ccchh~and \cccch~show that all three hydrocarbons are found until the border of the PDRs, despite the strong radiation fields ($G_0\sim 100-1000$). They moreover correlate well with each other, in agreement with trends reported in diffuse clouds (Lucas \& Liszt 2000). In the Horsehead nebula, the hydrocarbon large-scale distribution seems closely related to the emission of aromatic bands observed in the mid-IR, a result in contrast with the CO brightness distribution. A firm link between the small hydrocarbons and the aromatic band carriers however still requires high-resolution maps of this area (see Pety et al. 2003). Heavier species have been probed at positions of interest. They are mostly detected in the Horsehead nebula, with a noticeable detection of \csh~around 81.8\,GHz. To our knowledge, this is the first detection of this heavy carbon chain reported in a PDR. 

We have derived the hydrocarbon column densities and abundances relative to H$_2$ and \cch. On the average, the column densities of \cdo~and of the most abundant hydrocarbons are similar in all three PDRs, and comparable to results reported in other sources (e.g. Fuente et al. 1993, 2003). They exhibit representative ratios of order $\sim 25$, 1 and 15 for [\cdo]/[\cch], [$c$-\ccchh]/[\cccch] and [\cch]/[$c$-\ccchh], respectively. We have compared these values to column densities reported in the TMC-1 and L134N dark clouds, known as efficient carbon factories. Although H$_2$ column densities may be underestimated in the dark clouds, abundances relative to H$_2$ are similar up to chains and rings of 2-3 carbon atoms. The same trend is evidenced in carbon budgets derived for PDRs, diffuse clouds and dark clouds conditions. Ratios relative to \cch~however show lower values for the PDRs in comparisons with TMC-1, a very likely consequence of the different chemistries at work in respectively illuminated and well-shielded clouds. 

PDR model computations (Le Bourlot et al. 1993) using two independent gas-phase chemistry networks succeed in reproducing some of the observed abundances (\cdo, \cch) and column density ratios ([\cdo]/[\cch], [$c$-\ccchh]/[\cccch]). They however fail in producing the observed amount of $c$-\ccchh~and~\cccch~by an order of magnitude. This discrepancy may be to due to incorrect dissociation rates used in the networks, or to missing gas-phase reactions in the considered network. An additional formation process of hydrocarbon could also be at work. We propose to consider further the role of the AIB carriers in the chemistry as their erosion by the UV radiation may return small carbon molecules in the gas phase (see e.g. Allain et al. 1996, Scott et al. 1997, Le Page et al. 2001, 2003). New laboratory experiments (Joblin 2002) confirm that the photo-dissociation of PAHs can generate various kind of carbon clusters whose further dissociation leads to the release of small fragments, essentially C$_3$ and C$_2$. Further understanding and implementation of this mechanism in ISM models is needed in the future. More observational data, especially on heavier molecules, would also complete the scarce information on the carbon chemistry of PDRs.

\begin{appendix}
\section{Column density computations}
\label{appendix a}
The various column densities considered in the study (Sect.~\ref{coldens}) have been computed under distinct approximations, depending of the species and/or the source. We describe here the assumptions and details of their computation.

\subsection{C$^{18}$O}
\label{c18o}

We applied the method described in e.g. Abergel et al. (2003), consisting in using the optically thick emission lines of \dco~to infer the kinetic temperatures in all three sources. The values obtained at the CO emission peaks are indicated in Sect.~\ref{coldens}. Based on these temperatures, we used a LVG approximation and fitted, when applicable, the ($J$=1--0) and ($J$=2--1) transitions of \cdo~with pairs of [n$_{\rm H_2}$,$N$(\cdo)]. When only one transition was available (e.g. IC63), an additional assumption was made on the volume density n$_{\rm H_2}$.

\subsection{C$_2$H}
\label{c2h}

We took {\bf advantage} from the simultaneous detection of 4 components of the hyperfine structure of the N=1--0 transition of \cch~and applied the so-called HyperFine Structure method (HFS, see e.g.) featured in the CLASS reduction software. At each position, the fit returns the total transition opacity $\tau_{\rm tot}$, as well as the product $[J_\nu(T_{\rm ex})-J_\nu(T_{\rm bg})]\times \tau_{\rm tot}$ (with $T_{\rm bg}$ the cosmological background temperature). The method gives {\bf satisfactory} results provided the signal-to-noise ratio is sufficient, and the opacities are not too low ($\tau_{\rm tot} > 0.1$).

For optically thin or marginally optically thick lines, the column density is directly proportional to the integrated line intensity. For the excitation temperature ranges inferred in the three objects, $N$(\cch) is not sensitive to $T_{\rm ex}$, and we finally adopted the following expressions (in cm$^{-2}$):
\begin{eqnarray}
\label{eqn nc2h}
N({\rm C_2H}) = (7\pm1)\cdot 10^{13} \times W({\rm C_2H})/G(\tau)
\end{eqnarray}
where $G(\tau)=(1-{\rm e}^{-\tau^\prime})/\tau^\prime,~\tau^\prime = 0.679 \times \tau^{0.911}$, is a correction factor accounting for non-negligible opacity corrections (Wyrowski et al. 1999), and $\tau$ is the opacity of the main line at 87.316\,GHz.

\subsection{$c$-C$_3$H$_2$}
\label{c3h2}

Based on LVG simulations, we found that the $c$-\ccchh~intensities can be very sensitive to density variations in the conditions encountered in our targets ($1-5 \times 10^4$\,cm$^{-3}$, $T_{\rm kin}\simeq 20-50$\,K). As an illustration, considering the Horsehead case, the density uncertainties derived from the \cdo~analysis would lead to $N(c$-\ccch) variations larger of 50\%. To solve this problem, we used the good correlation observed between the brightness temperatures of $c$-\ccchh~and \cch~(e.g. Fig.~\ref{hh correl}). Since \cch~is much less affected by density variations, and considering the low optical depth of the molecules, this trend suggests homogeneous physical conditions in the clouds, although the excitation of each species strongly differs. We thus assumed a constant volume density and a representative kinetic temperatures to apply to the LVG simulations. Since our LVG code treats {\it ortho} and {\it para} forms separatly, we applied to the {\it ortho} $c$-\ccchh(2$_{1,2}$--1$_{0,1}$) transition a standard {\it o/p} ratio of 3.

\subsection{C$_4$H}
\label{c4h}

The low line intensities suggest very likely optically thin lines. Applying the same reasoning as for $c$-\ccchh, we performed LVG simulations on the \cccch(9--8) transition lines. Since the model does not account for the hyperfine structure of this transition, we built synthetic lines from the emission of all observed components. This combination uses the expected (at LTE) relative line intensity ratios $\alpha_N$, and writes:
\begin{eqnarray}
\label{c4h}
W_{\rm synth} = \frac{\Sigma\,\left(W_N/\alpha_N\right)}{N}
\end{eqnarray}
where $N$ is the n$^{\rm th}$ hyperfine component. In practice, the relative intensities of the (N=9--8, J=19/2--17/2, hereafter labelled 1) and (N=9--8, J=17/2--15/2, labelled 2) transitions (consisting both of two blended lines) are 0.53 and 0.47 respectively, so that $W_{\rm synth}\simeq W_1+W_2$. The outputs of the LVG simulations are given in Tab.~\ref{tab2}.

\subsection{Other species}
\label{other coldens}

The emissions exhibited by the rarer molecules introduced in Sect.~\ref{obs} ($l$-\ccch, $c$-\ccch, $l$-\ccchh, $l$-\cccch$_2$, \chhhcch, \csh~and HC$_3$N) have been analysed under LTE assumptions, using the densities and excitation temperatures inferred from the analyses described previously. In particular, for $c$-\ccch~and $l$-\ccch~we used the $T_{\rm ex}$ derived from \cch. For \csh, there are no cross sections available. We estimated them from those of HC$_3$N (Green \& Chapman 1978, -- see Cernicharo et al. 1999). Finally, for HC$_3$N, we assumed the $T_{\rm kin}$ values inferred in Sect.~\ref{c18o}. 
In case of non-detection, upper limits of the column densities are given (Tab.~\ref{tab2}).

\subsection{H$_2$ column densities}
\label{coldens h2}

Meaningful comparisons of the observed molecular column densities require an accurate estimate of the total H$_2$ column densities at each of the positions reported in Tab.~\ref{tab2}. We discuss here the methods used to derive as accurate as possible values for each of our targets.

\subsubsection{\roph}
\label{h2 roph}

In this source, we cannot use the extinction map of Wilking \& Lada (1983) as our positions lie outside their coverage. We nevertheless reproduced their approach, consisting in converting \cdo~column density into visual extinction, and applied the $N($H$_2)/A_{\rm V}$ calibration by Bohlin et al. (1978), which assumes a gas-to-dust ratio of 100. The choice of the the $A_{\rm V}/N({\rm C^{18}O})$ ratio is not straightforward as most reported values have been obtained in cold dark clouds where depletion onto grains affects the measurements and consequently increase this ratio (e.g. Frerking et al. 1982, Cernicharo et al. 1987). One can however recompute this ratio for warmer conditions assuming that all CO is in the gas phase. Using high-resolution infrared spectroscopy of H$_2$ and CO in NGC2024 IRS2, Kulesa et al. (2002) recently reported a new measurement of the CO abundance relative to H$_2$, [CO]\,$=3\times 10^{-4}$. With [O$^{16}$]/[O$^{18}$]=560 (Wilson \& Rood 1994) and using standard conversion numbers, one can infer:
\begin{eqnarray}
\label{coldens1}
A_{\rm V}/N({\rm C^{18}O}) = R \times 6.6 \times 10^{-16}~~{\rm mag\,cm^2}
\end{eqnarray}
where $R=A_{\rm V}/E(B-V)$ has reported values ranging between $\sim 3-5$ (Spitzer 1978, Ducati et al. 2003 and ref. therein). This yields:
\begin{eqnarray}
\label{coldens2}
A_{\rm V}/N({\rm C^{18}O}) = 2-3.3 \times 10^{-15}~~{\rm mag\,cm^2}
\end{eqnarray}
in very good agreement with the ratio proposed by Wilking \& Lada (1983). A zero point offset in $A_{\rm V}$ is however to be considered to take into account the fact that C$^{18}$O is photo-dissociated for low extinctions. We used an offset of 1.5\,mag, consistent with the calibrations by Frerking et al. (1982) and Cernicharo et al. (1987). We finally adopted the following conversion rule for \roph:
\begin{eqnarray}
\label{coldens3}
N({\rm H_2}) = (2.5\pm0.6) \cdot 10^{6}\times N({\rm C^{18}O}) +1.4\cdot 10^{21}~{\rm cm^{-2}}
\end{eqnarray}
At the peak \cdo~column density, this yields $N($H$_2)=1.1\times 10^{22}$\,cm$^{-2}$.

\subsubsection{Horsehead nebula}
\label{h2 hh}

In the Horsehead nebula, no extinction estimate is available at spatial resolution of order 30\arcsec. With a beam resolution of $\sim2$\arcmin, Kramer et al. (1996) report $A_{\rm V}$\,$\sim$\,3. This is consistent with rough estimates from the IRAS 100\,$\mu$m emission (4\arcmin~beam) but is very likely not adapted to the small-scale structure of this source. Applying the same approach as in \roph~implies extinctions of order 8--13\,mag. In order to cross-check these estimates, we have our observations of the continuum at 1.2\,mm (Fig.~\ref{hh maps}) and exploited them in the fashion described by Motte et al. (1998), with:
\begin{eqnarray}
\label{coldens motte}
N_{\rm H_2} = \frac{S_{\rm 1.2mm}}{\Omega_{\rm beam}\,\mu\,{\rm m_H}\,\kappa_{1.2}\,B_{1.2}(T_{\rm dust})}
\end{eqnarray}
where $S_{\rm mm}$ is the 1.2\,mm flux density in mJy per 11\arcsec~beam, $\Omega_{\rm beam}$ the beam solid angle, $\mu$ the mean molecular weight, m$_{\rm H}$ the mass of atomic hydrogen, $B_{1.2}(T)$ the Planck function at wavelength 1.2\,mm and temperature T, and $\kappa_{1.2}$ the dust mass opacity at 1.2\,mm. Such estimates are affected by the approximative knowledge of the dust properties at these wavelengths (e.g. Ossenkopf \& Henning 1994, Stepnik et al. 2003). With the additional assumption that $T_{\rm dust}$ is well represented by $T_{\rm kin}$ ($\simeq 40$ here), and adopting $\kappa_{1.2} = 3-6 \times 10^{-3}$\,g\,cm$^{-2}$, we derived from the observed peak flux of 56\,mJy/beam a peak extinction of 12--25\,mag. After smoothing to the 3\,mm beam, the corresponding extinction range reduces to 9--15\,mag, which translate into a column density of $N($H$_2)=1.1\pm0.3\times 10^{22}$\,cm$^{-2}$. This is significantly above any previous estimates (cf Sect.~\ref{presentation HH}), but in very good agreement with the independent calculation using \cdo~emission.

\subsubsection{IC63}
\label{h2 ic63}

In IC63, we simply used the value adopted by Jansen et al. (1994) $N($H$_2) = 5\pm2 \times 10^{21}$\,cm$^{-2}$, although it may be overestimated at positions away from the CO peak.

\end{appendix}

\begin{acknowledgements}
We would like to thank an anonymous referee whose comments helped to improve the paper and correct some numerics in the tables. We are also particularly grateful to J. Cernicharo for providing and helping us with his LVG code upgraded with several of the species presented in this paper. We are also grateful to C. Joblin for constructive discussions on the photo-dissociation of PAHs. 
\end{acknowledgements}

{}

\end{document}